\documentclass{article} % For LaTeX2e
\usepackage{iclr2027_conference,times}

\usepackage{amsmath,amsfonts,bm}

\def\eqref#1{equation~\ref{#1}}
\def\1{\bm{1}}

\DeclareMathAlphabet{\mathsfit}{\encodingdefault}{\sfdefault}{m}{sl}
\SetMathAlphabet{\mathsfit}{bold}{\encodingdefault}{\sfdefault}{bx}{n}

\usepackage{hyperref}
\usepackage{url}
\usepackage[table]{xcolor} % colors
\usepackage{amsmath}
\usepackage{algorithm}
\usepackage[noend]{algpseudocode}
\usepackage{xspace}
\usepackage{longtable}
\usepackage{booktabs}
\usepackage{array}
\usepackage{graphicx}
\usepackage{booktabs}
\newcolumntype{L}[1]{>{\raggedright\arraybackslash}p{#1}}
\usepackage{multirow}
\usepackage{capt-of}
\usepackage{tablefootnote}
\usepackage{wrapfig}
\usepackage{cleveref}
\usepackage{enumitem}

\usepackage{algorithm}
\usepackage{array,tabularx}
\usepackage{inconsolata}
\usepackage{microtype}
\usepackage{arydshln}
\usepackage{subcaption}
\usepackage{placeins}

\setlist[enumerate]{itemsep=2pt}
\setlist[itemize]{itemsep=0pt}

\definecolor{planaccent}{HTML}{245B78}
\definecolor{planband}{HTML}{EAF1F5}

\definecolor{rimecolor}{HTML}{0E939D}
\definecolor{poemscolor}{HTML}{543AAC}

\definecolor{alzero}{HTML}{48AE74}
\definecolor{alone}{HTML}{DF9138}
\definecolor{altwo}{HTML}{C32C2C}
\newcommand{\alcolor}[1]{\ifcase#1\color{alzero}\or\color{alone}\or\color{altwo}\fi}

\newcommand{\alz}{\textcolor{alzero}{AL0}\xspace}
\newcommand{\alo}{\textcolor{alone}{AL1}\xspace}
\newcommand{\alt}{\textcolor{altwo}{AL2}\xspace}

\newcommand{\namefont}{\fontencoding{T1}\fontfamily{qpl}\selectfont}

\newcommand{\datasetname}{{\namefont\color{rimecolor}\textsc{RIME}}\xspace}
\newcommand{\methodname}{{\namefont\color{poemscolor}\textsc{POEMS}}\xspace}

\definecolor{GPTColor}{HTML}{B88760}
\definecolor{GemmaColor}{HTML}{214E59}
\definecolor{GeminiColor}{HTML}{887AA3}
\definecolor{SFTColor}{HTML}{769C8C}

\newcommand{\gpt}{\textsc{\textcolor{GPTColor}{GPT-Audio-Mini}}\xspace}
\newcommand{\gemini}{\textsc{\textcolor{GeminiColor}{Gemini 3 Flash}}\xspace}
\newcommand{\gemma}{\textsc{\textcolor{GemmaColor}{Gemma 3n}}\xspace}
\newcommand{\gemmasft}{\textsc{\textcolor{SFTColor}{Gemma 3n SFT}}\xspace}

\newcommand{\planstep}[1]{\textcolor{planaccent}{{\bfseries $\hookrightarrow$ #1}}}
\newcolumntype{Y}{>{\raggedright\arraybackslash}X}

\title{\datasetname: Enabling Large-Scale Agentic \\ Music Post-Production}

\author{Noah Schaffer \& Nikhil Singh \\
Dartmouth College\\
\texttt{\{noah.schaffer.gr, nikhil.u.singh\}@dartmouth.edu} 
}

\iclrfinalcopy % Uncomment for camera-ready version, but NOT for submission.
\begin{document}

\maketitle

\begin{center}
\vspace{-2em}
\includegraphics[width=\linewidth]{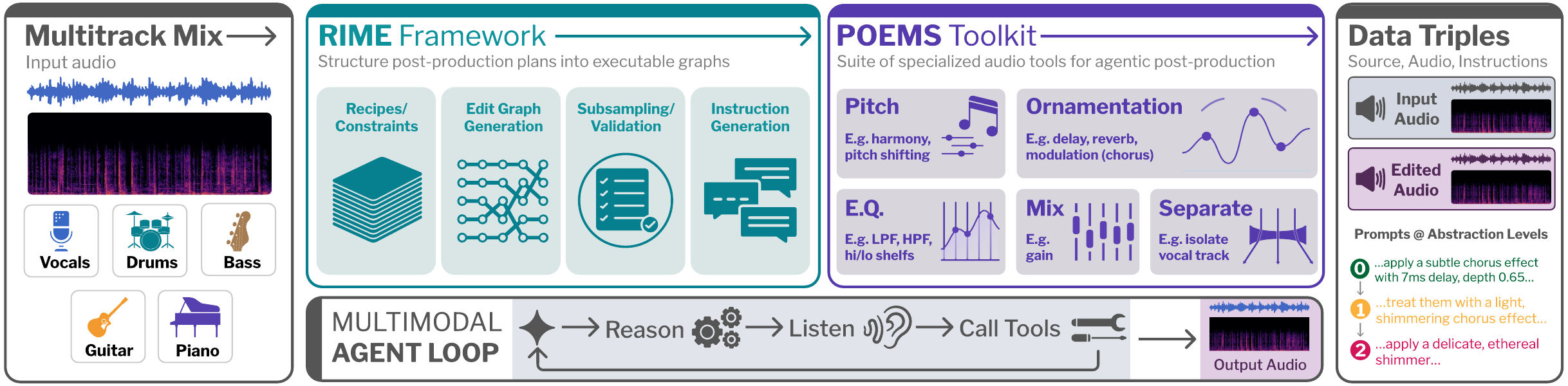}
\vspace{-1.5em}
\captionof{figure}{Agentic music post-production with \datasetname. Starting from a multitrack mix, \datasetname creates rule-based plans and calls \methodname tools to generate data triples of \texttt{(source audio, output audio, editing instruction)}. Downstream multimodal agents then use \methodname to implement an agentic loop in which they realize the provided instruction.}
\label{fig:banner}
\end{center}

\begin{abstract}
Almost every piece of recorded music you have ever heard was modified before it reached you; commercial releases rarely spring fully formed from a musician's mind. Despite the promise of music generation models for one-shot output, such fine-grained \textit{iterative refinement} workflows are a complementary problem, and largely out of their reach. There is also a gap for musicians: while they can express what they want to hear, not all have the facility with studio production tools to implement the complex set of actions needed to realize these intuitions. We formalize this task as \textit{agentic post-production}, wherein individual aspects of a track are targeted, refined, and combined into a final version. We argue the bottleneck is data: existing corpora do not reflect how realistic post-production chains map onto the vocabulary musicians and engineers actually use. We observe that there is a language for modifying recorded music that is dense, consistent, and learnable. To leverage this, we introduce the \textbf{{\namefont\color{rimecolor}R}ule-based {\namefont\color{rimecolor}I}nstructions for {\namefont\color{rimecolor}M}usic {\namefont\color{rimecolor}E}diting (\datasetname)} framework, which generates realistic paired edit-instruction data from any baseline music dataset grounded in canonical methods, design patterns, and constraints derived from real production workflows. \datasetname leverages \methodname, a new toolkit that combines stem separation, mixing, and common studio effects for use by multimodal agents. We use \methodname and \datasetname to generate 15,000 pairs of edit instructions and ground-truth audio, and use this data to evaluate existing multimodal LLMs as agents on this task, showing persistent challenges in current models' post-production capabilities. We also demonstrate \datasetname's ability to improve post-production agent performance via supervised fine-tuning using synthetic data generation. We see \datasetname as an early step towards iterative musical agents, collaborative systems that could transform music production much as interactive coding agents have reshaped software engineering.\footnote{We release the code at \href{https://github.com/sahaslab/RIME}{github.com/sahaslab/RIME} and a demo page at \href{https://www.sahaslab.com/rime}{sahaslab.com/rime}} 
\end{abstract}

\section{Introduction}
\label{intro}
There is a pervasive loop in the studio. When a producer listens to a recording they're working on, they will often hear something wrong, or something that could be better. They may isolate the offending stem, apply a chain of effects, blend it back in, and listen again. The musician being recorded may hear something else and make a request. The producer implements a fix, then listens again. Then again. Then again.

This loop has its own peculiar language. When a musician asks for something ``warmer'' or ``bigger,'' a seasoned engineer can translate it into a specific recipe (e.g. a high-shelf cut at 8 kHz, parallel compression on the drum bus, and a tempo-synced reverb send on the vocal) and run it. The translation between what music \textit{feels} like and the sequence of operations that produces that feeling is not, in fact, magic. It is a dense and surprisingly consistent vocabulary, built up over decades of studio practice.

Today's AI music systems are typically not part of this loop. Yet, we have good reason to desire them to be: it admits an iterative, interactive workflow that is musician-centered at its core. It is now reasonable to ask whether a software agent could join it, because these agents live in loops of their own: perception, action, reasoning, and tool use. Indeed, this would also be a strong testbed for such agents. Each iteration in this loop couples a perceptual judgment, an instruction context, a large assortment of potential tools, planning, and reasoning. What would it take, then, to bring an agent's loop into dialog with this \textit{post-production} loop?

We argue that what is needed is a general framework for modifying music, beyond the capabilities of one-shot generators. To this end, we introduce the \textbf{{\namefont\color{rimecolor}R}ule-based {\namefont\color{rimecolor}I}nstructions for {\namefont\color{rimecolor}M}usic {\namefont\color{rimecolor}E}diting (\datasetname)}  framework, which aims to generate large-scale music editing data that follows a process closer to what you would see in studio music production. \datasetname starts from individual mixes, creates edit ``recipes'' using a large vocabulary of common studio engineering rules, constraints, and design patterns, and executes a series of tool calls to isolate sources, apply edits, and create a final mix. This then creates supervision for agents to do the same. Additionally, \datasetname creates instructions at varying levels of abstraction to describe the edit chain for a given recipe, mimicking the variation in how precisely these intents are described in the real world. To create editing data, \datasetname leverages a newly built suite of tools that allow for {\namefont\color{poemscolor}\textbf{p}}itch, {\namefont\color{poemscolor}\textbf{o}}rnamentation (various effects), {\namefont\color{poemscolor}\textbf{e}}qualization, {\namefont\color{poemscolor}\textbf{m}}ixing, and source {\namefont\color{poemscolor}\textbf{s}}eparation (\methodname) edits, which we call \methodname.

\datasetname is also unique in that it can be extended to any dataset containing musical tracks, and especially designed for \textit{mixed} tracks (or even \textit{generated} ones, potentially, which normally come without individual instrument stems). We demonstrate using \datasetname in conjunction with state-of-the-art music understanding models to generate triplet edit data from untagged musical mixes. 

Overall, our contributions are:
\begin{enumerate}
    \item We formalize \textbf{agentic post-production} as an iterative task: starting from a mixed track and a production intent (an arbitrarily clear instruction), an agent targets individual elements, refines them, and recombines them into a finished mix
    \item We introduce \datasetname, \textbf{a scalable, reusable data generation framework} for transforming \textit{virtually any corpus of (contemporary Western) musical audio} into a structured set of (input, output, instruction) triples. \datasetname encodes recipes, motifs, ordering constraints, and pattern policies of studio post-production, in part calibrated from datasets of real production chains, and produces instructions at multiple levels of abstraction.
    \item To support this framework, we introduce \methodname, \textbf{an audio post-production toolkit} spanning 20+ operators exposed via Model Context Protocol for use by tool-calling agents.
    \item We run \textbf{a paired triplet benchmark} with $\approx$15,000 (input, output, instruction) triples generated with \datasetname on MusicCaps~\citep{agostinelli2023musiclm}, enabling evaluation across instrument families, edit categories, and abstraction levels.
    \item We conduct a \textbf{listener study} to validate that musically experienced users can reliably match edits and their corresponding instructions across abstraction levels.
    \item We develop the \methodname Agent Framework to design and evaluate \textbf{zero-shot agent} baseline multi-step tool-calling audio-language model agents, surfacing major limitations in current-generation models' abilities on this task, such as their ability to assemble the correct chain of operators vs. parameterize those operators correctly.
    \item We \textbf{fine-tune} an open-weight LLM orchestrator with \datasetname-synthesized data and demonstrate improvements over the zero-shot agent baselines.
\end{enumerate}

\section{Related Work}
\label{related}

Many previous text-prompted generative editing tasks have relied solely on paired instruction/audio data and audio-language embedding improvements to perform edits \citep{liang2024wavcraft, tsai2024audio, chu2025text2fx, zhang2024musicmagus, manor2024zero, yang2026melodia, lan2026declarative}. While this allows for certain edits without the need for supervised (input, output, text) data, the lack of ground-truth data makes these methods difficult to evaluate systematically. Several general audio editing frameworks leverage triplet data (input audio, edited audio, instruction); however, their datasets largely focus on more general edits such as addition, deletion, replacement (via text-to-audio synthesis), inpainting, and superresolution \citep{wang2023audit, liang2025audiomorphix, jia2025audioeditor, ellis2025recomposer}. InstructME \citep{han2024instructme} creates paired triplet edit data from musical variations, focusing on only a small subset of production edits such as remixing, adding, replacing, extracting, and removing. 

Agentic editing is a relatively unexplored area. SonicMaster \citep{melechovsky2025sonicmaster} uses an agent to repair five categories of degradations (EQ, Dynamics, Reverb, Amplitude, and Stereo) from audio, but does so using DDIM inversion and noise injection. LLM2Fx-Tools \citep{doh2025llm2fx} creates a benchmark from MedleyDB \citep{bittner2014medleydb, bittner2016medleydb}, randomly sampling musically realistic FX chains on single sources and uses a language model to create specific natural language instructions. It then fine-tunes a (text-only) LLM tool-calling agent to derive FX chains from instructions. Although this is the closest work to a post-production agent, it operates only on single stems, does not rely on an iterative listening workflow, and does not differentiate the multiple levels at which an edit may be described in natural language. 

Several existing frameworks have analyzed the mapping between abstract natural language and studio effects. SoxialFX~\citep{zheng2016socialfx} analyzed common abstract descriptors of three individual effects. Further work analyzed effect chain recommendation from natural language descriptors~\citep{stasis2017audio}. Recently, MixAssist~\citep{clemens2025mixassist} introduced a dataset capturing conversations between amateur and professional music producers during a mixing session. There is no existing work that has leveraged these findings to motivate actions taken by a tool-calling agent, nor have these language/FX mappings been extended to the full context of music post-production. 

In all, no existing benchmarks match the characteristics of real-world post-production loops described earlier. For this, such a benchmark would need to (i) apply edits to individual stems and introduce them back into a mix, (ii) be able to handle instructions at varying abstraction levels, and (iii) support an iterative, listen-and-tweak workflow. Here, we introduce exactly this suite of components.

\section{\datasetname: {\namefont\color{rimecolor}R}ule-based {\namefont\color{rimecolor}I}nstructions for {\namefont\color{rimecolor}M}usic {\namefont\color{rimecolor}E}diting}
\label{rime}
We assume that a user begins with an input audio clip $x$ with associated/extracted metadata $m(x) = (\mathcal{S}_x, g_x, t_x, \kappa_x)$ (set of available stems, genre tags, tempo, key). We seek to simulate many edit workflows to produce a dataset, outlined in \Cref{fig:banner} and explained below.

\begin{figure}[!htb]
    \centering
    \includegraphics[width=\linewidth]{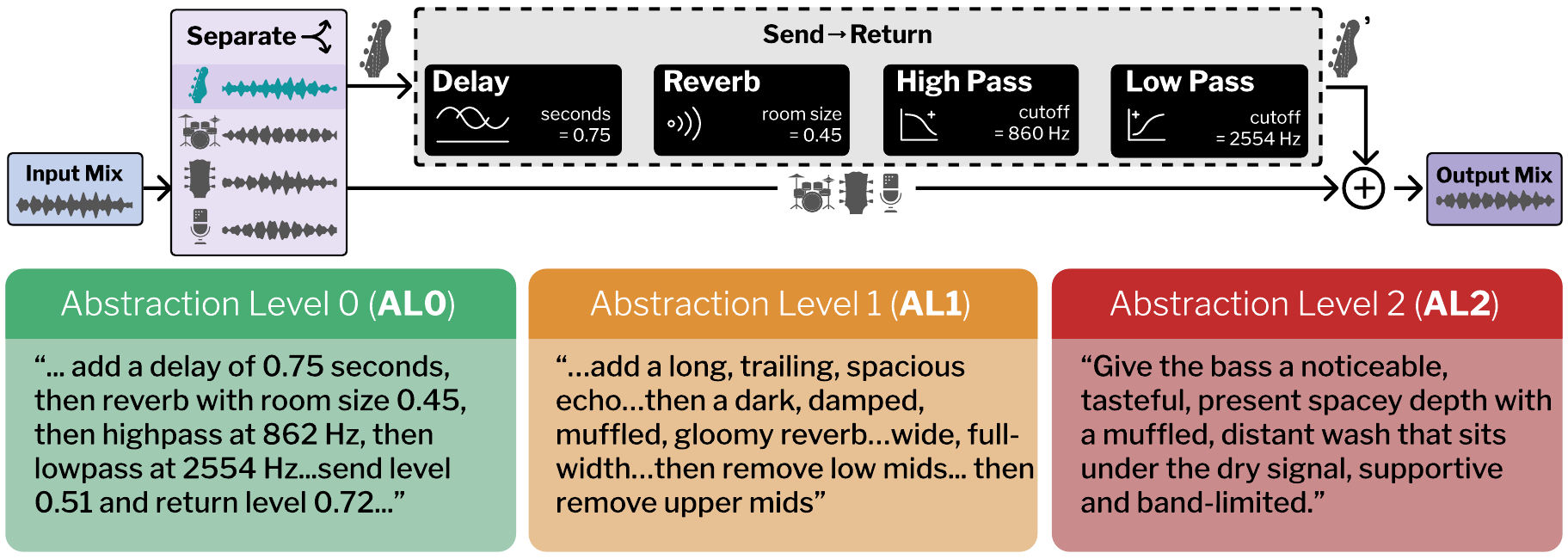}
    \vspace{-1.5em}
    \caption{A \datasetname edit graph instantiated from the \texttt{target\_\allowbreak space\_\allowbreak send\_\allowbreak synced} recipe on a \texttt{bass} stem. From this graph, we generate instructions at three distinct levels of abstraction.}
    \label{fig:chain}
\end{figure}

\subsection{The \methodname Toolkit}
\label{poems}

First, we need a suite of tools to enact a given edit workflow. Modern LLM agents can sometimes write code to create such tools, but we see no need to reinvent the wheel at runtime. Producers often have favored audio plug-ins, and integrating with these similarly requires the agent to use existing tools. At the same time, we need a stable, canonical, and agent-oriented set of tools for a robust evaluation. For this purpose, we introduce the Pitch, Ornamentation, Equalization, Mixing, and Separation (\methodname) toolkit, which defines an operator catalog $\mathcal{O}$ for end-to-end music post-production. Each operator $o \in \mathcal{O}$ has an associated parameter space $\Theta_o$.

To edit an individual source within a mixed track, we must first isolate it using {\namefont\color{poemscolor}source separation}. We use the 6-stem variant of Demucs~\citep{rouard2023hybrid} to separate $x = x_s + r_s$, where $x_s$ is the target stem and $r_s = x - x_s$ is the residual. This restricts editable stems to vocals, drums, bass, piano, and guitar (we attempted to use SAM Audio~\citep{shi2025sam}, but found it produced poor results in this setting). After processing, we {\namefont \color{poemscolor} mix} the edited stem back with its residual $\tilde{x} = \tilde{x}_s + r_s$.

We provide two {\namefont\color{poemscolor} pitch} edits for vocal stems: pitch correction (autotuning) and harmony generation. For both edits, we use estimated $f_0$ contour from CREPE~\citep{kim2018crepe}, S-KEY key detection~\citep{kong2025s}, quantization, and resynthesis with PSOLA~\citep{valbret1992voice} to preserve formant structure. We implement four {\namefont\color{poemscolor}ornamentation and equalization} effect families with Pedalboard~\citep{sobot_peter_2023_7817838}: time-based (delay, algorithmic reverb), modulation (chorus, phaser), dynamics (compressor, peak limiter, waveshaping distortion, noise gate, split-band de-esser), and equalization (high-pass, low-pass, high-shelf, low-shelf, and parametric peaking filters). In all, this gives us a substantial toolkit capable of enacting a wide variety of desired edits (list in \Cref{app:poems_tools}).

\subsection{\datasetname Components}

\paragraph{Recipes, Graphs, and Motifs}
The central abstraction \datasetname relies on is an \textit{edit graph} $G = (V, E)$, a directed acyclic graph whose nodes $v \in V = (o_v,\theta_v,s_v)$ each apply an audio operator $o_v \in \mathcal{O}$ with parameters $\theta_v \in \Theta_{o_v}$ to audio. This is similar to an FX chain in common audio production parlance, except that it allows parallel branches (via audio sends which copy audio off from the stream, and audio returns which re-mix modified versions of those copies into the original stream).

Hand-specifying enough graphs (e.g. thousands) to form a robust dataset would be infeasible. Rather, we need a way to \textit{generate} such graphs at scale. Fortunately, audio production follows many \textit{recipes}, defined here as symbolic templates for edit graphs with particular aesthetic goals (see example in \Cref{fig:chain}). Each recipe $r \in \mathcal{R}$ is specified using a schema $r = (B_r,\varphi_r,\lambda_r,H_r)$:
\begin{itemize}
    \item \texttt{bindings} $B_r$: (references to clip metadata resolved at plan-generation time, e.g. \textit{key})
    \item \texttt{when} $\varphi_r$ (eligibility predicate, e.g. \texttt{vocal\_harmony\_support} requires \textit{vocal stem} and \textit{key})
    \item \texttt{weight\_rules} $\lambda_r$ (multipliers to indicate how common certain recipes should be)
    \item \texttt{graph} $H_r$ (outline for applying the recipe, normally \texttt{SEPARATE} $\to$ \texttt{PROCESS} $\to$ \texttt{MIX})
\end{itemize}

\texttt{PROCESS} contains the recipe's actual processing sub-graph. \texttt{SEPARATE} and \texttt{MIX} scaffold it with stem isolation and residual recombination steps, so every recipe operates in a stem-aware fashion.

To share structure \textit{across} recipes, we factor commonly co-occurring sub-chains into reusable units we call \textit{motifs}. A motif is a named, typed sub-graph that can be expanded inline within any recipe that references it. For example, the \texttt{retro\_tilt\_eq} motif expands to \texttt{high\_shelf()} $\to$ \texttt{lowpass()}, capturing the joint high-frequency roll-off characteristic of a vintage sound.

\paragraph{Parameter Priors}
\label{methods:priors}
To generate plausible-sounding audio effects, we need to sample their parameters in such a way that we favor commonly used settings while allowing naturalistic variation across edits. We take an empirical approach to this by estimating prior sampling distributions from observed processor settings, which we call \textit{generation priors}. Each prior $\pi_o$ is a distribution over operator $o$'s argument space $\Theta_o$. Recipes specify which prior $\pi_o$ to use, so the same operator can use different priors in different recipes (e.g. to account for stylistic variation).

We collect processor settings from MixParams~\citep{clemens2025mixassist} and SocialFX~\citep{zheng2016socialfx}, using MixParams for equalization and compression, including separate vocal-compression priors, and SocialFX to supplement general compression thresholds and ratios. To balance coverage across sources and settings, we give contributing datasets equal weight, songs equal weight within MixParams, and distinct parameter values equal weight within SocialFX. We exclude controls with unspecified/unclear units (e.g. attack and release), and choose heuristic priors for these.

For each parameter $j$, we set heuristic bounds $\mathcal{D}_j = [a_j,b_j]$ to exclude settings we expect to produce excessively weak or extreme effects, and fit priors using observations within that domain. We then fit on $z = h_j(\theta_j) \in [0,1]$, applying a log transform before normalization for log-scaled parameters. Candidate distributions $f_m(z; \eta_m)$ include uniform, Gaussian, beta, and two-component Gaussian mixture models, plus power laws in original units for log-scaled parameters. We bound-constrain all candidates and fit by weighted maximum likelihood:

\begin{equation}
\widehat{\eta}_m = \arg\max_{\eta_m} \sum_i w_i\log f_m(z_i;\eta_m).
\end{equation}

Then, we select:
\begin{equation}
m^\star = \arg\min_m \left[ -2N_{\mathrm{eff}} \sum_i w_i\log f_m(z_i;\widehat{\eta}_m) + d_m\log N_{\mathrm{eff}}\right],
\end{equation}
where $d_m$ = number of fitted parameters and $N_{\mathrm{eff}}=1/\sum_g W_g^2$, with $W_g$ denoting total weight of one song or settings group. This is a BIC-like approximation to balance fit against complexity. We only include contributing sources with $\ge 30$ observations and at least three songs or distinct parameter settings groups to avoid overfitting to idiosyncratic examples. We use a custom web-based UI to have an audio engineer double-check parameter distributions by ear, described in \Cref{app:validation}.
 
\paragraph{Constraints}
\label{rime:constraints}
Plausible parameter values do not guarantee useful edits: effect order and recipe applicability also matter. \datasetname applies two classes of constraints when sampling edit graphs.

\textit{Chain Order Constraints.} We use the default order \texttt{EQ} $\prec$ \texttt{Dynamics} $\prec$ \texttt{Distortion} $\prec$ \texttt{Modulation} $\prec$ \texttt{Time-Based Effects} $\prec$ \texttt{Leveling} (see \Cref{app:rime_details} for specific references regarding effects ordering). Individual chains can specify a different order when needed; for example, ambiance recipes apply delay and reverb \textit{before} filtering the wet signal, while vocal harmony applies pitch shifting, gain adjustment, then filtering.

\textit{Pattern Policy Constraints.} We use policy weights to favor edits more appropriate for each clip and target instrument, such as reducing the weight of modulation on drums (very unusual) and increasing the weight of vocal compression (very common). Given clip metadata $m$ and slot assignment $b$, we combine the recipe-local weight $\lambda_r(m,b)$ with applicable policy multipliers $\lambda_q(m,b)$:
\begin{equation}
\omega = \lambda_r(m,b)\prod_q \lambda_q(m,b).
\end{equation}

An audio engineer with professional production credits reviewed, auditioned, and refined these rules alongside the recipes, motifs, and priors using our web-based UI (\Cref{app:validation}). We provide the full policy in \Cref{rime_constraints} and component details in \Cref{app:rime_details}.

\begin{algorithm}[!htb]
\caption{Recipe-based plan pool construction.}
\label{alg:graph_gen}
\begingroup
\small
\setlength{\tabcolsep}{6pt}
\renewcommand{\arraystretch}{1.35}

\noindent\textbf{Inputs:} annotated clips $\mathcal{D}$, recipes $\mathcal{R}$, motifs, parameter priors $\{\pi_o\}$, and $J$ draws per assignment.\par

\noindent\textbf{Output:} a list $\mathcal{P}$ of candidate graphs and selection weights. Initialize $\mathcal{P}\gets[\,]$.\par
\medskip

\noindent\begin{tabular}{@{}>{\centering\arraybackslash}p{2em}>{\raggedright\arraybackslash}p{\dimexpr\linewidth-2em-2\tabcolsep\relax}@{}}
\multicolumn{2}{@{}>{\columncolor{planband}[0pt][0pt]}l@{}}{\strut\hspace{6pt}\textcolor{planaccent}{\textbf{\textit{For each clip} $x\in\mathcal{D}$}}} \\

\planstep{1} & \textbf{Read clip annotations.} Obtain $m(x)$: available stem descriptions, genre, tempo, and key. \\

\multicolumn{2}{@{}>{\columncolor{planband}[0pt][0pt]}l@{}}{\strut\hspace{6pt}\textcolor{planaccent}{\textbf{\textit{For each recipe} $r\in\mathcal{R}$}}} \\

\planstep{2} & \textbf{Assign recipe slots.} Generate each allowed assignment $b$, such as a choice of target stem. Apply the following steps to each assignment. \\

\planstep{3} & \textbf{Check recipe requirements.} Skip the assignment if $\varphi_r(m,b)$ is false. \\

\planstep{4} & \textbf{Apply selection rules.} Compute the weight $\omega$ from recipe and global rules. Skip candidates that a rule rejects or gives $\omega\leq0$. \\

\multicolumn{2}{@{}>{\columncolor{planband}[0pt][0pt]}l@{}}{\strut\hspace{6pt}\textcolor{planaccent}{\textbf{\textit{Repeat steps 5--8 for $J$ parameter draws per surviving assignment}}}} \\

\planstep{5} & \textbf{Expand motifs.} Fill recipe template $H_r$ with assignment $b$ and replace motifs with constituent steps. \\

\planstep{6} & \textbf{Resolve arguments.} Use fixed values, clip annotations, and recipe's priors to get $G$. \\

\planstep{7} & \textbf{Validate the graph.} Check arguments and constraints (\S\ref{rime:constraints}); stop with error if either check fails. \\

\planstep{8} & \textbf{Add to candidate pool.} Append $(x,r,b,\textsc{Serialize}(G),\omega)$ to $\mathcal{P}$ for subsequent work. \\
\end{tabular}

\endgroup
\end{algorithm}

\subsection{Instruction Generation}

We combine the recipes, motifs, priors, and constraints above to construct a candidate pool $\mathcal{P}$ (\Cref{alg:graph_gen}). For each clip $x$, we resolve recipe bindings against $m(x)$ and generate $J$ parameter draws per eligible assignment. We retain each graph with weight $\omega$ for subsequent subsampling.

In post-production, there are varying degrees of granularity with which you can describe the same set of edits. A sound engineer may be able to describe an edit with extreme specificity (e.g. ``apply a chorus with 20 ms delay and 0.65 depth''), while a musician desiring a similar result may say e.g. ``apply a delicate shimmer.'' We account for this by generating instructions at three \textit{abstraction levels}:

\begin{enumerate}
\item[\textbf{\alz}] Describes the edit graph directly. All effect names, exact parameter values, and operator order are preserved in the generated instruction (e.g. language of professional mix engineers).
\item[\textbf{\alo}] Summarizes output of \alz. All effect names are preserved, but exact parameters and operator order are not. The distribution of parameter values is given as context to abstractly describe parameterization (e.g. reverb with a room size of 0.25 might be described as ``small'' while 0.9 might be described as ``vast'').
\item[\textbf{\alt}] Summarizes the output of \alo, focusing on overall character and magnitude of an edit chain. Does not directly specify effect names, parameter values, or operator order.
\end{enumerate}

All instructions are generated using NVIDIA Nemotron 3 Super 120B~\citep{blakeman2025nvidia} as a held-out model; more details in \Cref{prompts}. We also provide a UI for auditing these instructions against graphs, shown in \Cref{app:validation}.

\subsection{Data Validation}
\label{methods:valid}

As a final check, we seek to ensure instructions and recipes are \textit{audible}, \textit{stylistically appropriate}, and \textit{accurate} at every abstraction level. We therefore screen examples for four types of errors: \textbf{(1) abstraction-level violations:} instructions fail to follow the requirements for their abstraction level; \textbf{(2) inconsistencies:} instructions misinterpret the preceding level's text; \textbf{(3) stylistic inconsistencies:} instructions describe edits that conflict with the source material, using MusicCaps~\citep{agostinelli2023musiclm} captions as context, and \textbf{(4) parameter incongruity:} edits combine parameter values in ways that produce implausible processing. We use Qwen3 235B~\citep{yang2025qwen3} as an LLM judge and discard examples it flags under any criterion. We then rank the remaining examples by cosine similarity between source and edited audio using MERT~\citep{yizhi2023mert} embeddings; we remove the lowest-scoring $5\%$ to screen out extremely atypical edits, then keep the next 5,000 by audibility as our primary set. In \Cref{app:judge}, we provide full validation procedure details.

\section{Experiments}
\label{sec:experiments}

\paragraph{\methodname-AF: the \textbf{A}gent \textbf{F}ramework}

We introduce a multi-turn tool-calling framework for an agent to use \methodname to isolate a target stem, process it, and mix it back into the original track. The agent first listens to the input and proposes a tool-use plan based on the user's instruction. At each step, it listens to the current audio and may revise its plan before selecting the next tool or returning the result. For malformed tool calls, we allow $n$ recovery attempts before invoking the tool with default arguments (\Cref{app:poems_tools}); we set $n=1$. We allow at most 20 tool-call steps (excluding recovery attempts). If the agent does not return edited audio within this limit, we declare \textbf{agent failure} and use the unedited audio as evaluation outputs.

\paragraph{Generating Synthetic Data with \datasetname}
We use \datasetname to synthesize evaluation triplets \texttt{(input\_audio, edited\_audio, instruction)} starting from a corpus of full mixes with no available stem-level annotations. We use Music Flamingo \citep{ghosh2025music} to identify instruments in MusicCaps~\citep{agostinelli2023musiclm} tracks, and keep only tracks that contain at least 2 stems and have at least one stem in $[\text{vocals},\text{bass}, \text{drums},\text{guitar}]$, resulting in 2899 tracks. From these, we generate $79,905$ plans and subsample $25,000$ as described in \Cref{alg:graph_gen}. Before filtering, $1,757$ plans were excluded because Demucs failed to isolate the target stem. Using the LLM filtering described in \Cref{methods:valid}, we obtain  15,335 valid clips and then further subsample the top 5,000 using the MERT ranking described in the same section. 

For SFT training data, we sample 3,000 random instrument combinations, with priors derived from \citet{nunes2014like, pereira2023moisesdb}, generate 105,741 plans, subsample to 25,000, and then filter using the LLM judge, resulting in 20,098 plans. We then use coverage-constrained stratified random sampling, where we select 1,000 of these plans such that all recipes used in creating ground-truth edit data are represented in the training set. For each chosen plan, we generate instructions at all three abstraction levels, resulting in 15,000 unique evaluation and 3,000 unique SFT training \texttt{(source\_audio, output\_audio, editing\_instruction)} triples.

\paragraph{Supervised Fine-tuning}
\label{experiments:finetune}
We perform supervised fine-tuning on an agent using \gemma~\citep{team2024gemma} orchestration. We fine-tune \textit{only} the reasoning step and do not enforce any specific tool calls. We use low-rank adaptation \citep{hu2022lora} with a rank of $16$, $\alpha$ of $16$ and learning rate of $2 \times 10^{-4}$. We train for one epoch on the 3,000 data triplets from the training dataset. We keep the multi-turn tool-calling framework identical to the zero-shot agent for comparison. 

\paragraph{Evaluation Metrics}

We use two primary modes of evaluation: audio similarity and edit-graph similarity. We use Fr\'echet Audio Distance \citep{roblek2019fr,gui2024adapting} and Kernel Audio Distance \citep{chung2025kad} in the MERT \citep{yizhi2023mert} embedding space. We also introduce edit similarity ($\Delta \text{sim}_a$), which measures the similarity between the direction of change (in the MERT embedding space) of the agent's output vs. input and the direction of change of the ground-truth edited audio from \datasetname. For graphs, our primary metric is Graph F1, which is the arithmetic mean of four F1 scores on graph \textit{kind}, \textit{operator}, \textit{stem}, and \textit{edge}. In \Cref{app:aux_results}, we provide several additional evaluations to corroborate the primary metric results.

\paragraph{Human Evaluation}
To test whether instructions adequately describe audible edits across abstraction levels, we conduct a listener study using 30 \textit{original--edited audio} pairs, rating three instructions per pair: the matching instruction, a hard negative targeting the same instrument and often similar effects, and another distractor, rating the correspondence of each with the audible edit on a 1--5 scale.

\section{Results}
\label{results}

\subsection{Human listeners can identify edits across abstraction levels}
\begin{wrapfigure}[15]{l}{0.6\textwidth}
    \centering
    \includegraphics[width=\linewidth]{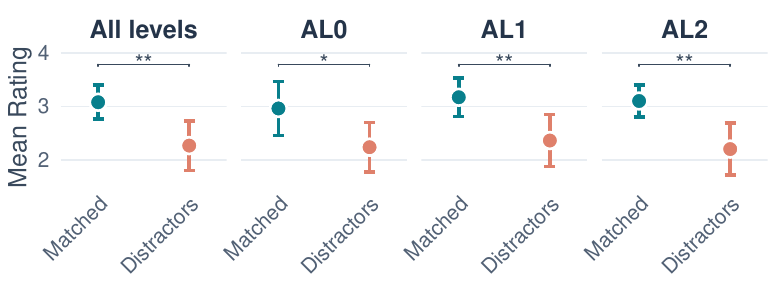}
    \vspace{-2em}
    \caption{Estimated marginal audio-correspondence ratings on a 1--5 scale for true edit instructions vs. distractors, adjusting for item differences. Listeners rated matching instructions above distractors across abstraction levels.}
    \label{fig:estimated-means}
    \vspace{-0.5em}
\end{wrapfigure}

We recruited 10 participants who all had music and audio backgrounds (via Prolific filters). Of the 10, 3 identified themselves as ``Mix engineer,'' 3 as ``Music producer,'' and 4 as ``Musician or performer.'' 3 had 10--19 years of experience, 4 had 5--9 years, 2 had 2--4 years, and 1 had less than 2 years. They described their listening setup as headphones (6), headphones and studio monitors (3), and studio monitors only (1). We collected 900 ratings across original--edited audio pairs (10 participants $\times$ 30 items $\times$ 3 ratings = 900 observations).

\begin{table}[!htb]
\centering
\small
\caption{Evaluation using pooled MERT clip embeddings. Graph F1 averages node-kind, stem, operator, edge F1. $\uparrow$ or $\downarrow$ indicate which direction is better.}
\label{tab:consolidated_results_full}
\begin{tabular}{@{}lllr rrrrr@{}}
\toprule
& & & & \multicolumn{5}{c}{Metrics} \\
& & Category & $N$ & FAD $\downarrow$ & FAD$_{\infty}$ $\downarrow$ & KAD $\downarrow$ & $\Delta \text{sim}_a$ $\uparrow$ & Graph F1 $\uparrow$ \\
\midrule
\multirow{4}{*}{\rotatebox[origin=c]{90}{\parbox{1.6cm}{\centering \gpt}}} & & \textit{Overall} & 15000 & 0.025 & 0.022 & 0.016 & 0.637 & 0.821 \\
& & \hspace{3mm} \alz & 5000 & 0.030 & 0.025 & 0.024 & 0.718 & 0.869 \\
& & \hspace{3mm} \alo & 5000 & 0.039 & 0.034 & 0.014 & 0.675 & 0.846 \\
& & \hspace{3mm} \alt & 5000 & 0.046 & 0.039 & 0.029 & 0.519 & 0.747 \\
\midrule
\multirow{4}{*}{\rotatebox[origin=c]{90}{\parbox{1.6cm}{\centering \gemini}}} & & \textit{Overall} & 15000 & 0.041 & 0.038 & 0.024 & 0.528 & 0.634 \\
& & \hspace{3mm} \alz & 5000 & 0.066 & 0.061 & 0.070 & 0.550 & 0.624 \\
& & \hspace{3mm} \alo & 5000 & 0.058 & 0.056 & 0.020 & 0.561 & 0.670 \\
& & \hspace{3mm} \alt & 5000 & 0.045 & 0.039 & 0.027 & 0.473 & 0.608 \\
\midrule
\multirow{4}{*}{\rotatebox[origin=c]{90}{\parbox{1.6cm}{\centering \gemma}}} & & \textit{Overall} & 15000 & 0.085 & 0.083 & 0.100 & 0.609 & 0.857 \\
& & \hspace{3mm} \alz & 5000 & 0.082 & 0.083 & 0.109 & 0.687 & 0.906 \\
& & \hspace{3mm} \alo & 5000 & 0.139 & 0.147 & 0.127 & 0.611 & 0.882 \\
& & \hspace{3mm} \alt & 5000 & 0.096 & 0.095 & 0.096 & 0.528 & 0.782 \\
\midrule
\multirow{4}{*}{\rotatebox[origin=c]{90}{\parbox{1.6cm}{\centering \gemmasft}}} & & \textit{Overall} & 15000 & 0.033 & 0.029 & 0.022 & 0.615 & 0.858 \\
& & \hspace{3mm} \alz & 5000 & 0.035 & 0.030 & 0.030 & 0.745 & 0.936 \\
& & \hspace{3mm} \alo & 5000 & 0.069 & 0.059 & 0.047 & 0.633 & 0.895 \\
& & \hspace{3mm} \alt & 5000 & 0.064 & 0.056 & 0.077 & 0.465 & 0.743 \\
\bottomrule
\end{tabular}

\end{table}

\begin{figure}[!htb]
\centering
\begin{minipage}{\linewidth}
    \phantomsubcaption\label{fig:ladder_continuous_by_abstraction}
    \phantomsubcaption\label{fig:parameter_tolerance_by_abstraction}
    \includegraphics[width=\linewidth]{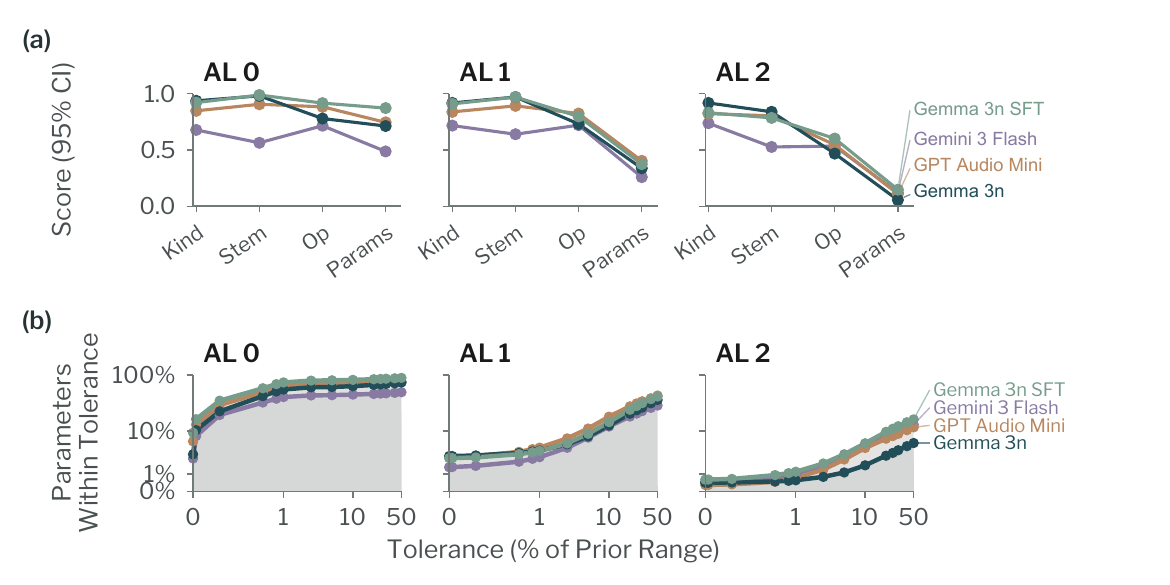}
\end{minipage}
\vspace{-1.5em}
\caption{Graph and parameter recovery across abstraction levels.
(a) Kinds, stems, and operators use F1; Params uses $1-$normalized MAE (b) Parameter accuracy as numeric tolerance increases.}
\vspace{-1em}
\label{fig:graph_parameter_recovery}
\end{figure}

In \Cref{fig:estimated-means}, we plot estimated marginal means from a regression predicting rating as a function of description role (the true matched description vs. distractor) interacted with abstraction level with source-audio (item) fixed effects and participant-clustered standard errors. We find that human listeners rated the true edit description as \textbf{significantly more matched to the audio} across all abstraction levels ($p=0.002$ for \alo, \alt, and $p=0.02$ for \alz). The mean rating is higher for the true description compared with both distractors at all levels of abstraction (all $p<0.04$ after Benjamini-Hochberg correction across 6 tests, except  \alz Distractor 1 $p = 0.12$). Additionally, the mean rating was higher for matched than for both distractors for all 30 items. Together, this provides strong evidence that the summarization and abstraction pipelines preserve necessary information pertaining to specific edits.

\subsection{Post-Production Agents Struggle in Zero-Shot Settings}

Across 15,000 attempts per model (5,000 target edits $\times$ 3 abstraction levels), \gpt produces 13,636 successful outputs (90.9\%), \gemini 13,563 (90.4\%), and \gemma 14,880 (99.2\%). Most zero-shot failures (95.9\%) reach the step limit without completing the edit. Others include invalid responses. We treat failed runs as no-ops with the input audio as output and an empty predicted edit graph.

Results are shown in \Cref{tab:consolidated_results_full}. In summary: \gpt has the highest target-similarity increase ($0.637$ vs. $0.609$ for \gemma and $0.528$ for \gemini), and the best FAD/FAD$_\infty$/KAD. \gemma wins on Graph F1 ($0.857$ vs. $0.821$ for \gpt and $0.634$ for \gemini), but also has worst FAD ($0.085$). \gemma completes more requests and recovers more graph structure than \gemini, yet produces a less similar output \textit{distribution}.

\paragraph{Abstraction challenges parameter recovery}
Edit similarity decreases as instructions move from explicit parameter values (\alz) toward evocative descriptions (\alt). \gpt falls from $0.718 \to 0.519$, and \gemma from $0.687 \to 0.528$. \gemini improves slightly from \alz to \alo ($0.550 \to 0.561$), then falls to $0.473$ at \alt. \gpt leads among zero-shot agents at \alz and \alo. \gemma is slightly ahead at \alt.

\begin{wrapfigure}{l}{0.4\linewidth}
\centering
\includegraphics[width=\linewidth]{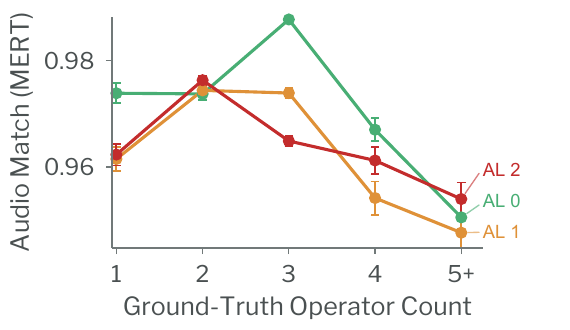}
\vspace{-1.5em}
\caption{Avg. MERT output--target similarity by ground-truth operator count. Error bars show 95\% CIs.}
\label{fig:complexity_stratification_overall}
\vspace{-2em}
\end{wrapfigure}

Parameter recovery is the clearest decline (\Cref{fig:ladder_continuous_by_abstraction}): $0.746 \to 0.403 \to 0.109$ for \gpt, $0.713 \to 0.337 \to 0.055$ for \gemma, and $0.487 \to 0.258 \to 0.118$ for \gemini. \Cref{fig:parameter_tolerance_by_abstraction} shows that this gap persists across accuracy thresholds; relaxing tolerance improves accuracy, but all models have a large gap between \alz $\to$ \alo $\to$ \alt; at 25\% of the prior range, zero-shot parameter accuracy is $20.9$--$31.7\%$ at \alo and $3.6$--$8.7\%$ at \alt.

\paragraph{Graph complexity also matters}
At \alt, operator F1 falls from $0.879$ for 1-operation edits $\to 0.442$ for 5+-operation edits for \gpt. Corresponding drops are $0.825 \to 0.451$ for \gemini and $0.764 \to 0.369$ for \gemma (\Cref{fig:complexity_stratification_overall}). Explicit instructions reduce this penalty: \gpt and \gemma have operator F1 $> 0.9$ for long graphs at \alz. These results identify two linked difficulties for zero-shot agents: inferring settings from abstract descriptions and recovering longer effect chains when operations are not explicitly specified.

\begin{wraptable}[18]{r}{0.4\linewidth}
    \vspace{-2em}
    \centering
    \small
    \begin{tabular}{@{}lrrr@{}}
\toprule
& \multicolumn{3}{c}{Accuracy @ $\tau$} \\
\cmidrule(l){2-4}
Model & $1/127$ & $10\%$ & $25\%$ \\
\midrule
\multicolumn{4}{@{}l}{\textit{AL0}} \\
\gpt & 59.0 & 69.9 & 72.5 \\
\gemini & 38.4 & 45.3 & 47.1 \\
\gemma & 51.4 & 63.5 & 68.2 \\
\gemmasft & 69.0 & 81.9 & 84.9 \\
\addlinespace
\multicolumn{4}{@{}l}{\textit{AL1}} \\
\gpt & 4.3 & 18.3 & 31.7 \\
\gemini & 2.6 & 12.1 & 20.9 \\
\gemma & 3.9 & 12.7 & 23.9 \\
\gemmasft & 3.7 & 14.6 & 28.5 \\
\addlinespace
\multicolumn{4}{@{}l}{\textit{AL2}} \\
\gpt & 0.6 & 4.6 & 8.0 \\
\gemini & 0.9 & 4.9 & 8.7 \\
\gemma & 0.5 & 1.8 & 3.6 \\
\gemmasft & 1.1 & 5.7 & 11.0 \\
\bottomrule
\end{tabular}

    \caption{Target-parameter accuracy (\%) at three tolerances $\tau$.}
    \label{tab:parameter_tolerance_by_abstraction}
    \vspace{-6em}
\end{wraptable}

\subsection{Simple plan SFT improves parameter recovery}
\gemmasft has the highest completion rate (14,953/15,000, 99.7\%). At \alz, it beats all other models in edit similarity, Graph F1, and parameter score. At \alo, it beats base \gemma on all main metrics, but \gpt is better at edit similarity and distributional audio metrics. At \alt, improvements are most clear in parameter and operator recovery, with other metrics mixed.

We test parameter recovery (\Cref{tab:parameter_tolerance_by_abstraction})  using a weighted model-by-abstraction linear regression with target-edit fixed effects and cluster-robust standard errors on a source clip. SFT reduces normalized MAE at every level (all $p<0.001$). Thus, even instruction-to-graph fine-tuning improves performance across abstraction levels.

\section{Conclusion}
The dominant paradigm in music ML is one-shot generation. The dominant paradigm in studio practice is iteration. Reconciling these competing visions requires data and infrastructure, and our work offers a scalable first step on this front. A first step always has limitations: \datasetname relies on audited heuristics where existing production data is missing, edits are limited to separable stems and a fixed operator pool, external baselines do not yet exist, and our audio metrics are only proxies for creative usefulness. Our benchmark also evaluates single instructions rather than multi-turn exchanges with a musician. We invite the community to build on this work to make AI-enhanced music production more like a session than a single take.

\bibliography{iclr2027_conference}
\bibliographystyle{iclr2027_conference}

\appendix
\crefalias{section}{appendix}
\crefalias{subsection}{appendix}

\section{\methodname Toolkit Details}
\label{app:poems_tools}

\subsection{\methodname Tool List}

\Cref{tab:poems-tools} shows a complete list of \methodname tools with descriptions and parameters.

\begingroup
\renewcommand{\arraystretch}{1.3}
\setlength{\tabcolsep}{4pt}
\newcommand{\poemsarg}[2]{\texttt{#1}\,=\,#2}
\centering
\begin{small}
\begin{longtable}{@{}L{2.9cm} L{2.3cm} L{3.2cm} L{4.7cm}@{}}
\caption{The Pitch, Ornamentation, Equalization, Mixing, and Separation tools of the \methodname toolkit. All tools except \texttt{mix\_stem\_with\_residual} and \texttt{separate} also take \texttt{audio} and \texttt{sr}. Default agent fallback values are listed next to parameters.}
\label{tab:poems-tools}\\
\toprule
\textbf{Tool name} & \textbf{Tool type} & \textbf{Description} & \textbf{Tool arguments (with default values)} \\
\midrule
\endfirsthead

\toprule
\textbf{Tool name} & \textbf{Tool type} & \textbf{Description} & \textbf{Tool arguments (with default values)} \\
\midrule
\endhead

\midrule
\multicolumn{4}{r@{}}{\textit{Continued on next page}}\\
\endfoot

\bottomrule
\endlastfoot

% --- Ornamentation ---
\texttt{apply\_chorus}     & Ornamentation & Chorus modulation effect. &
  \poemsarg{rate\_hz}{1.5}, \poemsarg{depth}{0.35}, \poemsarg{centre\_delay\_ms}{8.0}, \poemsarg{feedback}{0.15}, \poemsarg{mix}{0.35} \\
\texttt{apply\_phaser}     & Ornamentation & Phaser modulation effect. &
  \poemsarg{rate\_hz}{0.8}, \poemsarg{depth}{0.4}, \poemsarg{centre\_frequency\_hz}{1200.0}, \poemsarg{feedback}{0.2}, \poemsarg{mix}{0.35} \\
\texttt{apply\_distortion} & Ornamentation & Waveshaping distortion / saturation. &
  \poemsarg{drive\_db}{18.0} \\
\texttt{apply\_reverb}     & Ornamentation & Algorithmic reverb. &
  \poemsarg{room\_size}{0.65}, \poemsarg{damping}{0.45}, \poemsarg{wet\_level}{0.25}, \poemsarg{dry\_level}{0.75}, \poemsarg{width}{1.0}, \poemsarg{freeze\_mode}{0.0} \\
\texttt{apply\_delay}      & Ornamentation & Delay / echo with feedback. &
  \poemsarg{delay\_seconds}{0.25}, \poemsarg{feedback}{0.3}, \poemsarg{mix}{0.35} \\
\texttt{apply\_compressor} & Ornamentation & Dynamic-range compressor. &
  \poemsarg{threshold\_db}{$-$18.0}, \poemsarg{ratio}{3.0}, \poemsarg{attack\_ms}{10.0}, \poemsarg{release\_ms}{100.0} \\
\texttt{apply\_limiter}    & Ornamentation & Peak limiter. &
  \poemsarg{threshold\_db}{$-$1.0}, \poemsarg{release\_ms}{100.0} \\
\texttt{apply\_noisegate}  & Ornamentation & Noise gate. &
  \poemsarg{threshold\_db}{$-$45.0}, \poemsarg{ratio}{2.0}, \poemsarg{attack\_ms}{10.0}, \poemsarg{release\_ms}{150.0} \\
\texttt{apply\_deesser}    & Ornamentation & Split-band de-esser with relative-threshold detection. &
  \poemsarg{ess\_low\_hz}{4000.0}, \poemsarg{ess\_high\_hz}{10000.0}, \poemsarg{threshold\_db}{$-$24.0}, \poemsarg{ratio}{4.0}, \poemsarg{attack\_ms}{10.0}, \poemsarg{release\_ms}{100.0}, \poemsarg{relative\_threshold\_db}{$-$12.0}, \poemsarg{max\_reduction\_db}{9.0}, \poemsarg{filter\_order}{4} \\
\addlinespace[6pt]
% --- EQs ---
\texttt{apply\_highpass}   & EQ & Highpass filter. &
  \poemsarg{cutoff\_frequency\_hz}{120.0} \\
\texttt{apply\_lowpass}    & EQ & Lowpass filter. &
  \poemsarg{cutoff\_frequency\_hz}{8000.0} \\
\texttt{apply\_highshelf}  & EQ & High-shelf filter. &
  \poemsarg{cutoff\_frequency\_hz}{5000.0}, \poemsarg{gain\_db}{2.0}, \poemsarg{q}{0.707} \\
\texttt{apply\_lowshelf}   & EQ & Low-shelf filter. &
  \poemsarg{cutoff\_frequency\_hz}{120.0}, \poemsarg{gain\_db}{2.0}, \poemsarg{q}{0.707} \\
\texttt{apply\_peakfilter} & EQ & Parametric peaking EQ band. &
  \poemsarg{cutoff\_frequency\_hz}{2500.0}, \poemsarg{gain\_db}{2.0}, \poemsarg{q}{1.0} \\
\addlinespace[6pt]
% --- Mixing utilities ---
\texttt{apply\_gain}       & Mixing & Apply linear gain in decibels. &
  \poemsarg{gain\_db}{3.0} \\
\texttt{normalize\_peak}   & Mixing & Scale audio so its peak reaches a target level. &
  \poemsarg{target\_peak}{0.95} \\
\texttt{apply\_fade\_in\_out} & Mixing & Linear fade-in and fade-out at edit boundaries. &
  \poemsarg{fade\_in\_seconds}{0.02}, \poemsarg{fade\_out\_seconds}{0.02} \\
\texttt{apply\_pan}        & Mixing & Constant-power stereo panning. &
  \poemsarg{pan}{0.0} \\
\texttt{mix\_stem\_\allowbreak with\_residual} & Mixing & Sum a separated stem with its residual. &
  \texttt{stem}, \texttt{residual} \\
\addlinespace[6pt]
% --- Pitch ---
\texttt{autotune}          & Pitch & Quantize vocal pitch to a (detected or supplied) key, resynthesized via PSOLA. &
  \texttt{hcqt}, \texttt{chromanet}, \texttt{crop\_fn}, \texttt{device}, \texttt{key}, \texttt{mode} \\
\texttt{generate\_harmony} & Pitch & Generate a key-quantized harmony shifted by a given interval. &
  \texttt{hcqt}, \texttt{chromanet}, \texttt{crop\_fn}, \texttt{device}, \poemsarg{semitones}{7.0}, \texttt{key}, \texttt{mode} \\
\texttt{apply\_pitch\_shift} & Pitch & Constant pitch shift in semitones. &
  \poemsarg{semitones}{3.0} \\
\addlinespace[6pt]
% --- Separation ---
\texttt{separate}          & Separation & Separate a target stem and residual using either a SAM-Audio or a Demucs model. &
  \texttt{model}, \texttt{processor}, \texttt{device}, \texttt{audio}, \texttt{description}, \texttt{sample\_rate} \\
\end{longtable}
\end{small}
\endgroup

\section{\datasetname Details}
\label{app:rime_details}

\subsection{Recipe Catalog}
\label{app:recipes}

All \datasetname recipes share a common four-field schema: \texttt{bindings} (metadata
references resolved at plan-generation time), \texttt{when} (a predicate that
controls eligibility), \texttt{weight\_rules} (optional prior-shaping multipliers),
and \texttt{graph} (a declarative three-step edit graph of the form
\textsc{Separate} $\to$ \textsc{Process} $\to$ \textsc{Mix}).
\Cref{tab:recipe-summary} provides an overview of all \datasetname recipes.

We use the following references to inform effect-ordering heuristics. \footnote{Effects chaining references: \href{https://articles.boss.info/the-ultimate-guide-to-guitar-effects-pedal-order-and-signal-chain/}{1},\href{https://www.izotope.com/en/learn/signal-chain-order-of-operations}{2},\href{https://www.iconcollective.edu/mixing-effects-chain-order}{3}}

\begin{table}[h]
\centering
\small
\caption{Core authored \datasetname recipes, with categories, tags, target families, and base weights.}
\label{tab:recipe-summary}
\resizebox{\textwidth}{!}{
\begin{tabular}{llllc}
\toprule
\textbf{Recipe ID} & \textbf{Category} & \textbf{Tags} & \textbf{Target Family} & \textbf{Base Weight} \\
\midrule
\texttt{target\_space\_send\_synced}\tablefootnote{Reverb/space references: \href{https://www.musicguymixing.com/abbey-road-reverb-trick/}{1}, \href{https://moosicentertainment.com/2025/05/19/unlocking-clarity-the-abbey-road-reverb-trick-for-cleaner-mixes/}{2}, \href{https://flypaper.soundfly.com/produce/the-abbey-road-trick-how-to-eq-reverb-sends-to-free-up-space-in-a-mix/}{3}, \href{https://www.youtube.com/watch?v=L2qMXWP2UDo}{4},\href{https://www.audio-issues.com/music-mixing/this-simple-technique-makes-your-reverb-sit-perfectly-in-the-mix/}{5}} & Time-based   & delay, reverb, send/return & non-drum, non-other & 0.80 \\
\texttt{target\_space\_send\_free}   & Time-based   & delay, reverb, send/return & non-drum, non-other & 0.35 \\
\texttt{parallel\_drum\_compression}\tablefootnote{Parallel compression references: \href{https://www.musicguymixing.com/parallel-compression-on-drums/}{1}} & Dynamics     & compression, send/return   & drums               & 0.95 \\
\texttt{light\_drum\_distortion}     & Color        & distortion                 & drums               & 0.70 \\
\texttt{make\_target\_louder}        & Level        & gain                       & non-other           & 0.75 \\
\texttt{make\_target\_quieter}       & Level        & gain                       & non-other           & 0.75 \\
\texttt{vocal\_control\_compression}\tablefootnote{Vocal compression references: \href{https://www.musicguymixing.com/compressor-settings-for-vocals/}{1}, \href{https://www.sonarworks.com/blog/learn/vocal-compression-get-your-vocals-sounding-great}{2},\href{https://mastering.com/vocal-compression-how-to-compress-vocals/}{ 3},\href{https://www.reddit.com/r/audioengineering/comments/5uqf8b/almost_every_vocal_mixing_tutorial_says_to_put/}{4}, \href{https://unison.audio/vocal-compression/}{5}} & Dynamics     & compression                & vocals              & 0.90 \\
\texttt{vocal\_harmony\_support}\tablefootnote{Harmony references: \href{https://www.sageaudio.com/articles/how-to-mix-harmony-vocals}{1},\href{https://www.izotope.com/en/learn/ways-to-texturize-background-vocals}{2}}     & Pitch        & harmony, send/return       & vocals              & 0.45 \\
\texttt{retro\_tilt}\tablefootnote{Retro Tilt/EQ references: \href{https://www.youtube.com/watch?v=4kzcgQ229Rg}{1}, \href{https://abletunes.com/blog/eq-cheat-sheet/}{2},\href{https://www.reddit.com/r/diypedals/comments/1dsiuw3/high_shelf_eq/}{3}, \href{https://www.songstuff.com/recording/article/eq-frequencies/}{4}}                 & Color        & EQ (shelf, pass)           & non-drum, non-other & 0.35 \\
\texttt{modulation\_texture}\tablefootnote{Modulation references: \href{https://www.avid.com/resource-center/chorus-effect}{1},\href{https://www.reddit.com/r/guitarpedals/comments/18j2n6a/typical_depth_range_on_chorusflanger_pedals/}{2}, \href{https://static.roland.com/manuals/jd-08_reference/eng/17812188.html}{3}}         & Modulation   & chorus                     & non-drum, non-other & 0.30 \\
\texttt{hum\_cleanup} \tablefootnote{Hum cleanup references: \href{https://physics.stackexchange.com/questions/22552/strategies-against-50-hz-mains-hum-on-detector-signals}{1},\href{https://elfquake.wordpress.com/2017/12/29/revisiting-hum-filters/}{2},\href{https://www.reddit.com/r/DSP/comments/6cz1lp/removing_mains_hum_which_overlaps_with_desirable/}{3},\href{https://s3.amazonaws.com/izotopedownloads/docs/rx-plug-in-pack/De-hum.html}{4}, \href{https://www.digitimer.com/mains-interference-is-there-an-alternative-to-a-50-60hz-notch-filter/}{5}}               & Cleanup      & EQ (highpass)              & mixture-level       & 0.95 \\
\texttt{vocal\_sibilance\_cleanup}\tablefootnote{De-Essing references:  \href{https://www.izotope.com/en/learn/the-dos-and-donts-of-de-essing}{1}, \href{https://cymatics.fm/blogs/production/how-to-eq-vocals}{2},\href{https://www.reddit.com/r/audioengineering/comments/1b7i9es/de_essing_and_favorable_high_frequencies/}{3}}                & Cleanup      & EQ (peak cut)        & vocals       & 1.00 \\
\texttt{restore\_buried\_target} & Cleanup & gain & non-other & 0.85 \\
\texttt{tame\_overpowering\_target} & Cleanup & gain & non-other & 0.85 \\
\texttt{synthetic\_hum\_cleanup} & Cleanup & EQ (highpass) & mixture-level & 0.80 \\
\texttt{synthetic\_rumble\_cleanup} & Cleanup & EQ (highpass) & mixture-level & 0.80 \\
\texttt{control\_compression} & Dynamics & compression & non-vocal, non-other & 1.00 \\
\texttt{target\_highpass} & EQ & EQ (highpass) & non-other & 1.00 \\
\texttt{target\_lowpass} & EQ & EQ (lowpass) & non-other & 1.00 \\
\texttt{target\_highshelf} & EQ & EQ (high shelf) & non-other & 1.00 \\
\texttt{target\_lowshelf} & EQ & EQ (low shelf) & non-other & 1.00 \\
\texttt{target\_highpass\_compression} & Dynamics & EQ (highpass), compression & non-other & 1.00 \\
\texttt{target\_lowpass\_compression} & Dynamics & EQ (lowpass), compression & non-other & 1.00 \\
\texttt{target\_highshelf\_compression} & Dynamics & EQ (high shelf), compression & non-other & 1.00 \\
\texttt{target\_lowshelf\_compression} & Dynamics & EQ (low shelf), compression & non-other & 1.00 \\
\texttt{target\_eq\_limiter} & Dynamics & EQ (shelf, pass), limiter & non-other & 1.00 \\
\texttt{target\_chorus\_send} & Modulation & chorus, send/return & non-other & 1.00 \\
\texttt{target\_delay\_insert} & Time-based & delay & non-drum, non-other & 1.00 \\
\texttt{target\_delay\_send} & Time-based & delay, send/return & non-drum, non-other & 1.00 \\
\texttt{target\_reverb\_insert} & Time-based & reverb & non-other & 1.00 \\
\texttt{target\_reverb\_send} & Time-based & reverb, send/return & non-other & 1.00 \\
\texttt{target\_eq\_delay\_send} & Time-based & EQ (shelf, pass), delay, send/return & non-drum, non-other & 1.00 \\
\texttt{target\_eq\_reverb\_send} & Time-based & EQ (shelf, pass), reverb, send/return & non-other & 1.00 \\
\texttt{target\_compression\_delay\_send} & Time-based & compression, delay, send/return & non-drum, non-other & 1.00 \\
\texttt{target\_compression\_reverb\_send} & Time-based & compression, reverb, send/return & non-other & 1.00 \\
\bottomrule
\end{tabular}
}
\end{table}

\subsection{Motifs}
\label{rime_motifs}

\Cref{tab:motifs} highlights the reusable edit chains (``motifs'') for certain recipes. We describe both the types of effects used as well as the real-world application for each motif. 

\begin{table}[h!]
    \centering
    \small
    \caption{Reusable \datasetname motifs, their tags, and intended editing roles.}
    \label{tab:motifs}
    \resizebox{\textwidth}{!}{
    \begin{tabular}{lll}
\toprule
\textbf{Motif ID} & \textbf{Tags} & \textbf{Real-world Editing Role} \\
\midrule
\texttt{band\_limited\_space\_send\_synced} 

& delay, reverb, EQ pass filters 
& tempo-synced filtered ambience send \\

\texttt{band\_limited\_space\_send\_free}   
 
& delay, reverb, EQ pass filters 
& free-time filtered ambience send \\

\texttt{parallel\_compression\_bus}          
 
& compression 
& compressed parallel drum bus \\

\texttt{vocal\_control\_compression}         
  
& compression 
& serial vocal leveling compression \\

\texttt{retro\_tilt\_eq}                     
     
& EQ shelf, EQ pass filter 
& vintage-style high-frequency rolloff \\

\texttt{modulation\_texture}                 
 
& chorus 
& width and movement texture \\
\texttt{control\_compression} & compression & serial dynamics control for non-vocal stems \\
\texttt{target\_highpass} & EQ pass filter (highpass) & low-end cleanup / thinning \\
\texttt{target\_lowpass} & EQ pass filter (lowpass) & top-end rolloff / darkening \\
\texttt{target\_highshelf} & EQ shelf (high) & broad top-end boost or cut \\
\texttt{target\_lowshelf} & EQ shelf (low) & broad low-end boost or cut \\
\texttt{target\_highpass\_compression} & EQ pass filter (highpass), compression & low-end cleanup before leveling \\
\texttt{target\_lowpass\_compression} & EQ pass filter (lowpass), compression & top-end rolloff before leveling \\
\texttt{target\_highshelf\_compression} & EQ shelf (high), compression & top-end tone shaping before leveling \\
\texttt{target\_lowshelf\_compression} & EQ shelf (low), compression & low-end tone shaping before leveling \\
\texttt{target\_eq\_limiter} & EQ shelf, EQ pass filter, limiter & vintage tilt with peak control \\
\texttt{target\_chorus\_send} & chorus & fully wet chorus return for send routing \\
\texttt{target\_delay\_send} & delay & fully wet echo return for send routing \\
\texttt{target\_reverb\_send} & reverb & fully wet ambience return for send routing \\
\texttt{target\_delay\_insert} & delay & blended echo on the stem itself \\
\texttt{target\_reverb\_insert} & reverb & blended ambience on the stem itself \\
\bottomrule
\end{tabular}
    }
\end{table}

\subsection{Recipe Constraints}
\label{rime_constraints}

\begin{figure}[!htb]
    \centering
    \includegraphics[width=\linewidth]{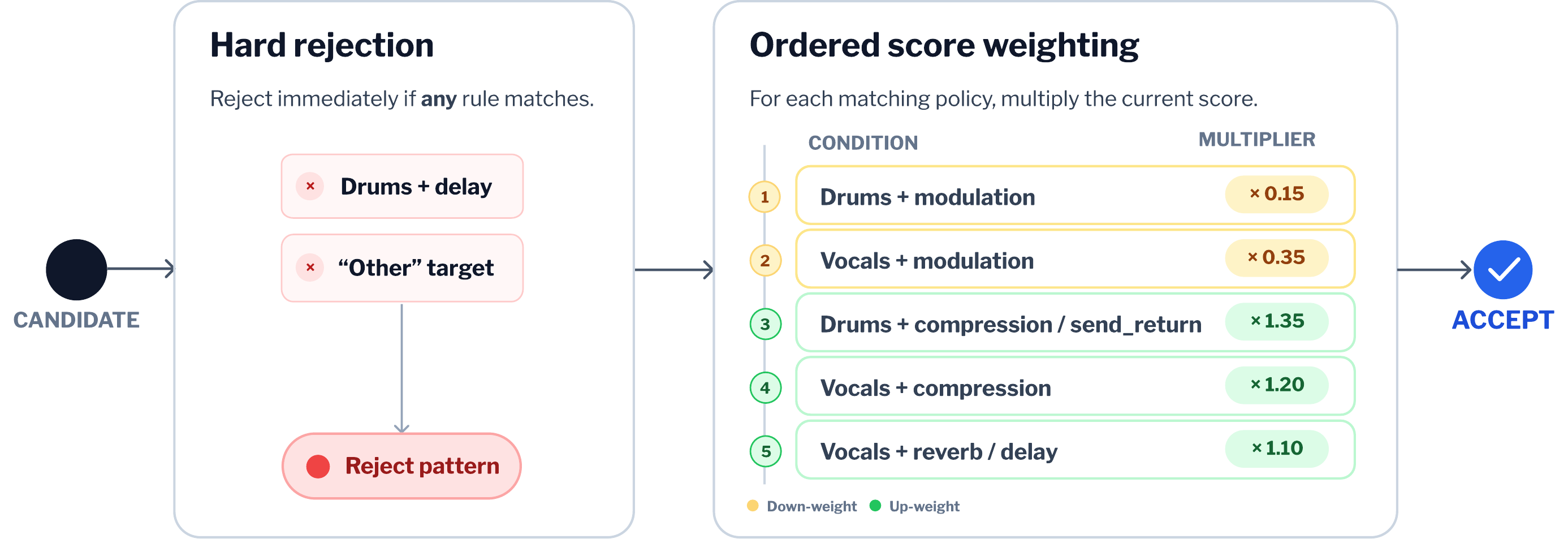}
    \caption{Pattern policy constraints in \datasetname. Pattern constraints control how edit recipe candidates are ranked during edit graph construction.}
    \label{fig:rime_constraints}
\end{figure}

\Cref{fig:rime_constraints} highlights how we use policy constraints to determine weighting for possible \datasetname recipes. Note that two policies are immediately rejected: applying delay on drums and applying any edits on the ``other'' source from Demucs. 

\subsection{Parameter Priors}
\label{app:priors}
\begin{figure}[!htb]
    \centering
    \includegraphics[width=\linewidth]{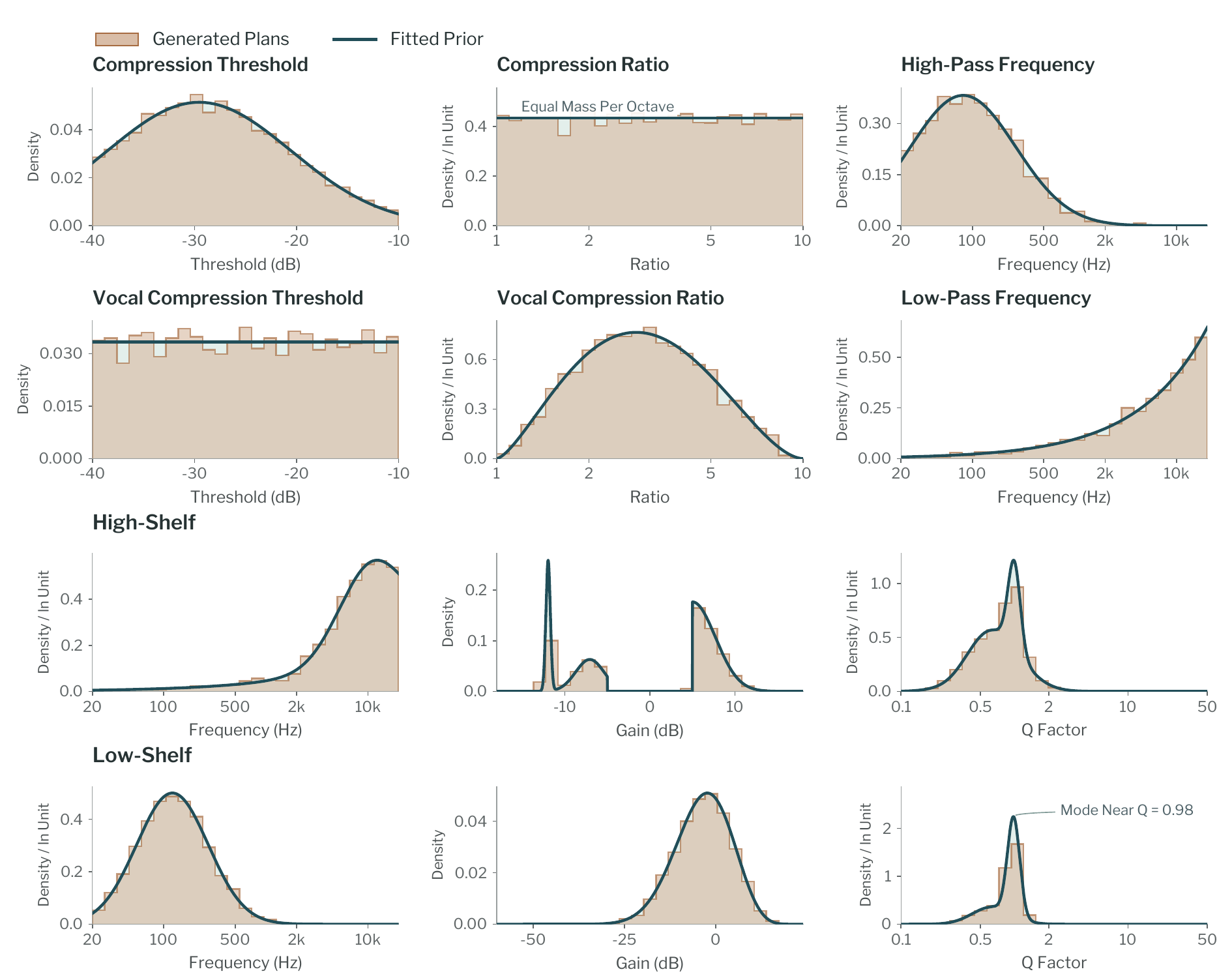}
    \caption{Continuous parameters with fitted priors. Histograms show samples from 25,000 subsampled candidate plans before filtering/ranking; curves show configured prior densities.}
    \label{fig:musiccaps_subsampled_sampling_density}
\end{figure}

\begin{figure}[!htb]
        \centering
    \includegraphics[width=\linewidth]{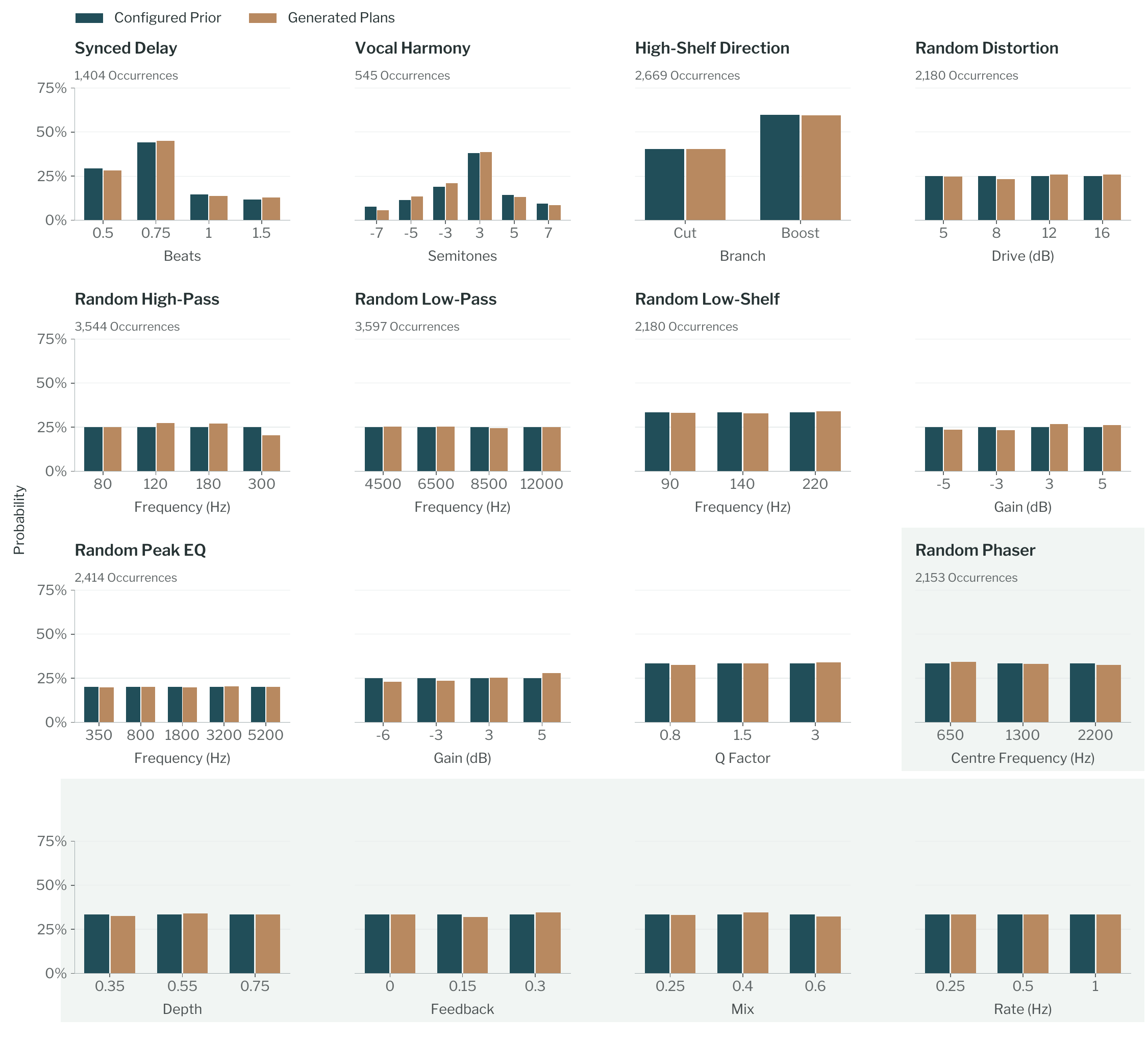}
    \caption{Discrete parameter priors. Bars compare configured choice probabilities with empirical frequencies in the 25,000 subsampled candidate plans.}
    \label{fig:musiccaps_subsampled_discrete_priors}
\end{figure}

\Cref{tab:generation-priors} shows the parameter priors for \datasetname tools. This combines both heuristically authored and audited priors with those fitted to external data as described in \Cref{methods:priors}. We plot both the continuous (\Cref{fig:musiccaps_subsampled_sampling_density}) and discrete (\Cref{fig:musiccaps_subsampled_discrete_priors}) fitted and sample prior distributions for each parameter in \datasetname (before filtering and ranking).

\begin{figure}[!htb]
    \centering
    \includegraphics[width=\linewidth]{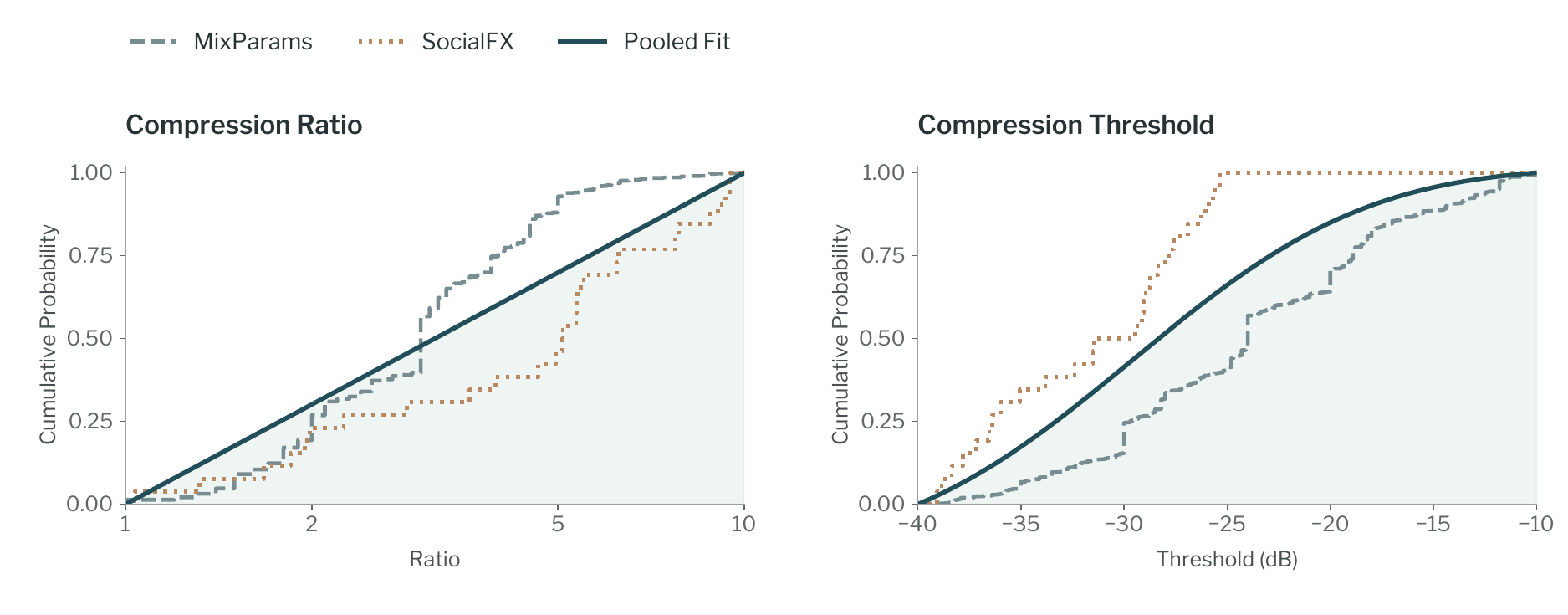}
    \caption{Empirical prior CDFs from our two source datasets for parameters where they both contribute, and the resulting fits (in blue).}
    \label{fig:source_comparisons}
\end{figure}

\begin{figure}[!htb]
    \centering
    \includegraphics[width=\linewidth]{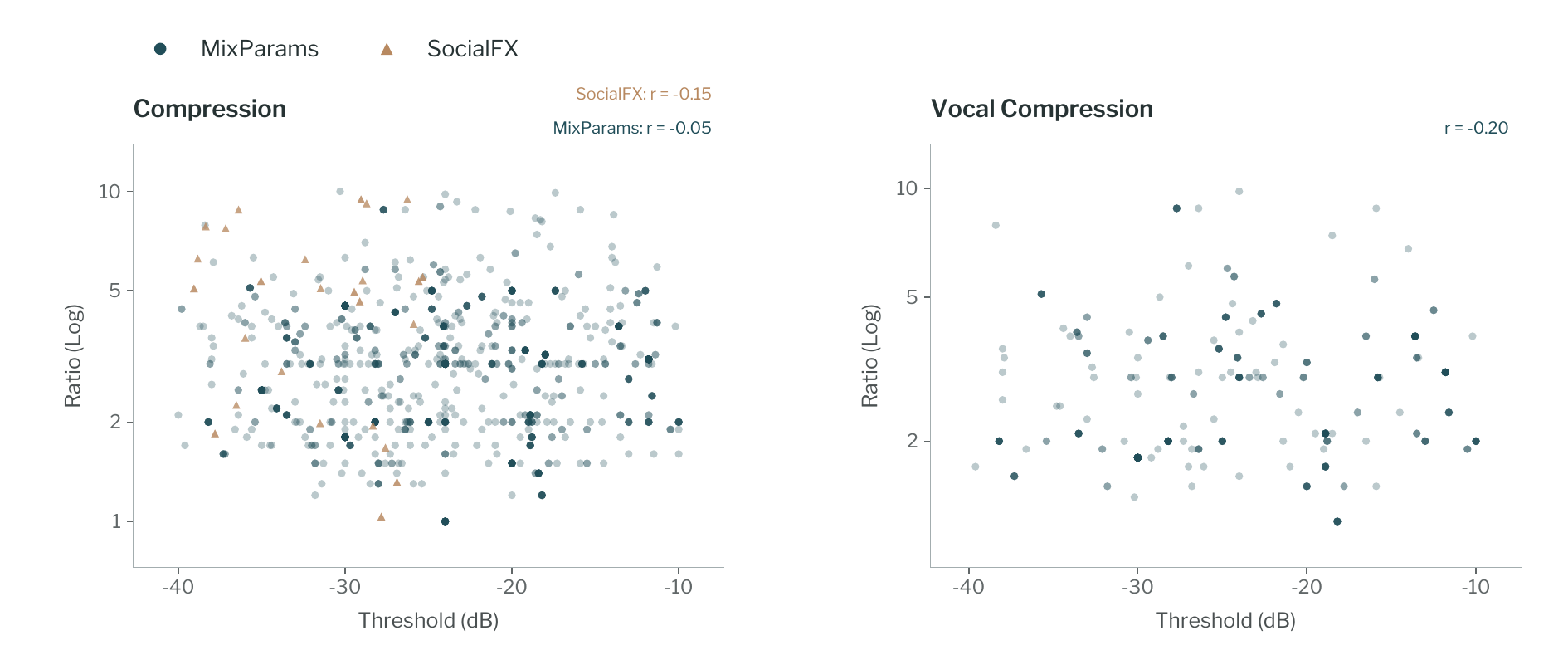}
    \caption{Compressor threshold vs. ratio, paired within the same source record and effect instance. Panels separate general and vocal compression; vocal observations are a subset of the general data. Labels report weighted Pearson correlations in the plotted coordinates, with ratio on a log scale.}
    \label{fig:source_relationships_compression}
\end{figure}

\begin{figure}[!htb]
    \centering
    \includegraphics[width=\linewidth]{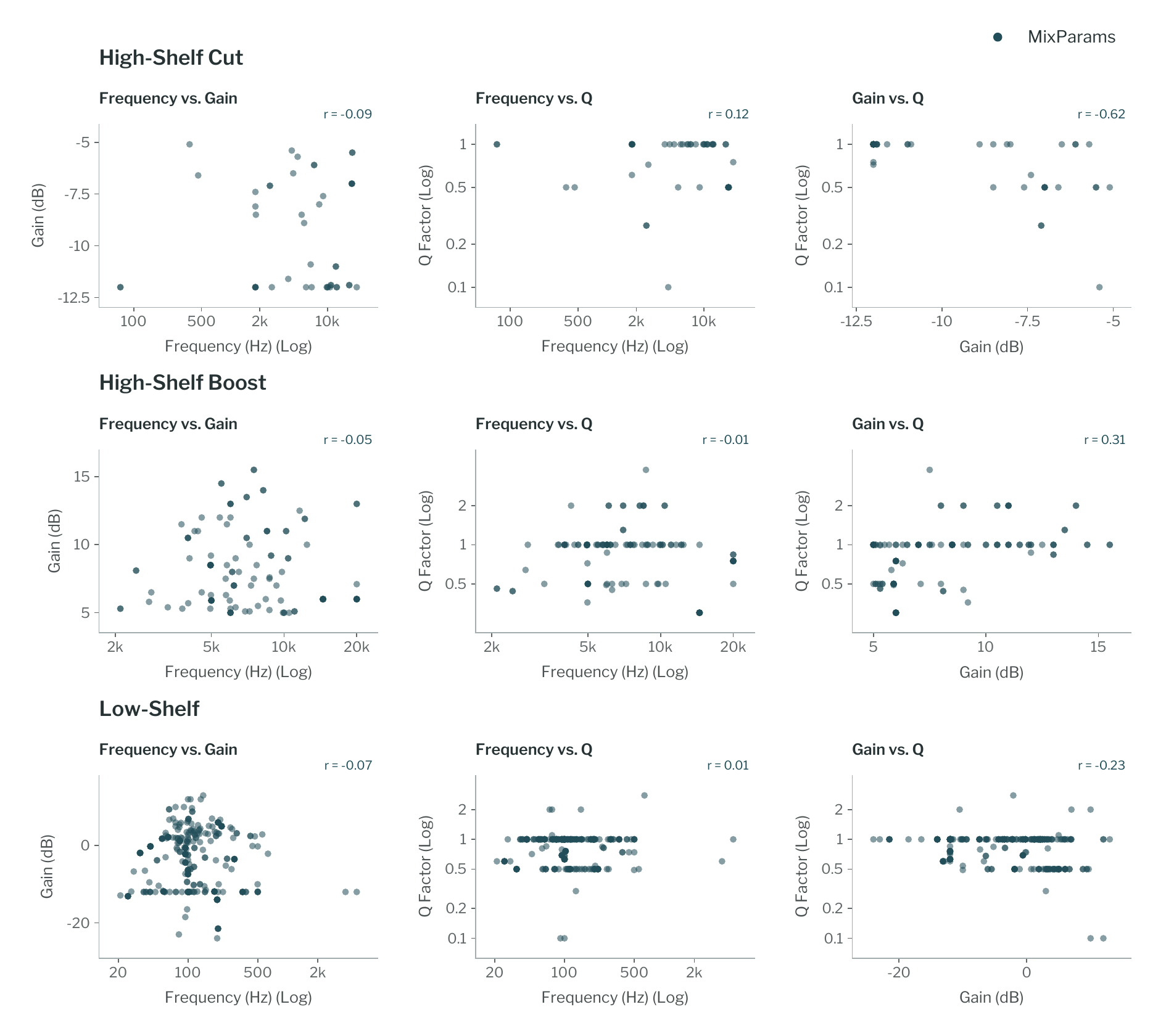}
    \caption{Within-effect EQ parameter relationships, with high-shelf boosts and cuts separated. Labels report weighted Pearson correlations (frequency and Q use log scales, while gain uses a linear dB scale). The high-shelf cut gain--Q association is stronger than the other associations ($r=-0.62$).}
    \label{fig:source_relationships_equalization}
\end{figure}

We provide plots of specific parameters from MixParams \citep{clemens2025mixassist} and SoxialFX \citep{zheng2016socialfx} as additional insight into parameter prior selection. In \Cref{fig:source_comparisons} we compare the two sources and pooled fits for compressor threshold and ratio. \Cref{fig:source_relationships_compression,fig:source_relationships_equalization} show within-effect parameter associations. Compressor associations are generally weak, with stronger associations confined to a few EQ pairs (e.g. high-shelf cut gain versus Q, $r=-0.62$). We thus make a simplifying independence assumption in our sampling scheme.

\begin{figure}[!htb]
    \centering
    \includegraphics[width=\linewidth]{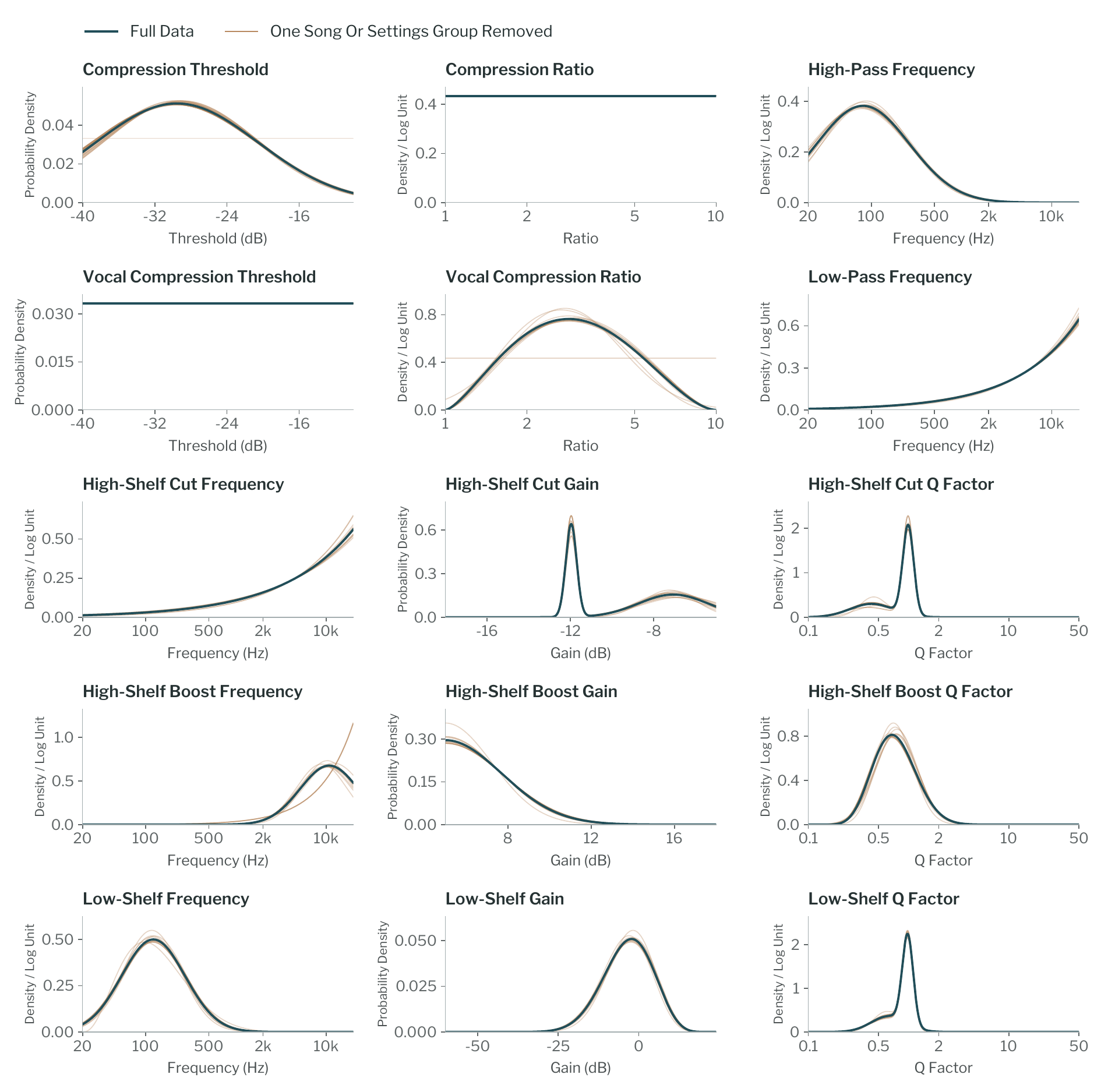}
    \caption{Sensitivity of fitted densities to removing one MixParams song or one SocialFX group with identical complete effect settings. Each reduced dataset is refitted using the same weighted maximum-likelihood fitting and family-selection procedure, with parameter bounds and filtering rules fixed. Distributions are largely stable.}
    \label{fig:source_stability_density}
\end{figure}

Finally, in \Cref{fig:source_stability_density}, we remove one MixParams song or one group of identical SocialFX presets at a time and refit the distributions using the same procedure. The fitted densities remain generally stable, suggesting that our priors are not driven by individual songs or repeated presets.
 
\begingroup
\footnotesize
\setlength{\tabcolsep}{3pt}
\renewcommand{\arraystretch}{1.22}
\begin{longtable}{@{}>{\raggedright\arraybackslash}p{\dimexpr0.24\linewidth-6pt\relax}>{\raggedleft\arraybackslash}p{\dimexpr0.17\linewidth-6pt\relax}>{\raggedright\arraybackslash}p{\dimexpr0.42\linewidth-6pt\relax}>{\raggedright\arraybackslash}p{\dimexpr0.19\linewidth-6pt\relax}@{}}
\caption{Complete FX generation priors. We sample parameters independently within each parameter set, conditional on branch for high-shelf settings. M denotes \href{https://huggingface.co/datasets/mclemcrew/MixParams/tree/df15a107cf86ed46793df859daf43d5f24b142fd}{MixParams}; S denotes \href{https://huggingface.co/datasets/seungheondoh/socialfx-original/tree/d7e266cac1526ae7db52ea16c744595a05b9df27}{SocialFX}. Parentheses show available record counts. General compression gives M and S equal total weight. $H$ denotes ranges or discrete choices we heuristically specified and audited by listening. We also choose the support bounds for every fitted prior heuristically. Counts can overlap across rows and high-shelf branches.}
\label{tab:generation-priors}\\
\toprule
Parameter & Support $[a, b]$ & Distribution & Source ($n$) \\
\midrule
\endfirsthead
\multicolumn{4}{@{}l}{\tablename\ \thetable\ (continued)} \\
\toprule
Parameter & Support & Distribution & Source ($n$) \\
\midrule
\endhead
\midrule
\multicolumn{4}{r@{}}{Continued on next page} \\
\endfoot
\bottomrule
\endlastfoot

\multicolumn{4}{@{}l}{\textbf{General compression}} \\*
% dynamics.compression.settings.parameters.threshold_db
Threshold (dB) & $[-40,-10]$ & $\mathcal{N}(-29.5419,9.0051)$ & M (1004), S (114) \\
% dynamics.compression.settings.parameters.ratio
Ratio & $[1,10]$ & $L(a, b)$ & M (1004), S (114) \\
\multicolumn{4}{@{}l}{\textit{Attack and release reuse the vocal-compression priors below.}} \\
\addlinespace
\multicolumn{4}{@{}l}{\textbf{Vocal compression}} \\*
% vocals.compression.threshold_db
Threshold (dB) & $[-40,-10]$ & $U(a, b)$ & M (353) \\
% vocals.compression.ratio
Ratio & $[1,10]$ & $B_{\log}(2.5088,2.8050)$ & M (353) \\
% vocals.compression.attack_ms
Attack (ms) & $[5,10]$ & $L(a, b)$ & $H$ \\
% vocals.compression.release_ms
Release (ms) & $[40,80]$ & $L(a, b)$ & $H$ \\
\addlinespace
\multicolumn{4}{@{}l}{\textbf{Equalization}} \\*
% tone.equalization.highpass.cutoff_frequency_hz
High-pass cutoff (Hz) & $[20,20000]$ & $\mathcal{N}_{\log}(4.3955,1.1836)$ & M (790) \\
% tone.equalization.lowpass.cutoff_frequency_hz
Low-pass cutoff (Hz) & $[20,20000]$ & $P(-0.3605)$ & M (124) \\
% tone.equalization.lowshelf.settings.parameters.cutoff_frequency_hz
Low-shelf cutoff (Hz) & $[20,20000]$ & $\mathcal{N}_{\log}(4.8001,0.8079)$ & M (287) \\
% tone.equalization.lowshelf.settings.parameters.gain_db
Low-shelf gain (dB) & $[-60,24]$ & $B(17.0721,8.3069)$ & M (287) \\
% tone.equalization.lowshelf.settings.parameters.q
Low-shelf Q & $[0.1,50]$ & $0.3626\,\mathcal{N}_{\log}(-0.4565,0.4015)$\newline ${}+0.6374\,\mathcal{N}_{\log}(-0.0193,0.1243)$ & M (287) \\
% tone.equalization.shelf_q
Q for heuristic shelf recipes & $[0.5,1]$ & $L(a, b)$ & $H$ \\
\addlinespace
\multicolumn{4}{@{}l}{\textbf{High-shelf branch and conditional parameters}} \\*
% tone.equalization.highshelf.settings.components: branch probabilities
Branch & Cut, boost & $\Pr(\mathrm{cut})=0.4032$;\newline $\Pr(\mathrm{boost})=0.5968$ & M (167) \\
% tone.equalization.highshelf.settings.components[cut].distribution.parameters.cutoff_frequency_hz
Cut: cutoff (Hz) & $[20,20000]$ & $P(-0.4502)$ & M (46) \\
% tone.equalization.highshelf.settings.components[cut].distribution.parameters.gain_db
Cut: gain (dB) & $[-18,-5]$ & $0.4172\,\mathcal{N}(-11.9597,0.2600)$\newline ${}+0.5828\,\mathcal{N}(-7.0590,1.6674)$ & M (46) \\
% tone.equalization.highshelf.settings.components[cut].distribution.parameters.q
Cut: Q & $[0.1,50]$ & $0.3748\,\mathcal{N}_{\log}(-0.8488,0.4977)$\newline ${}+0.6252\,\mathcal{N}_{\log}(-0.0083,0.1243)$ & M (46) \\
% tone.equalization.highshelf.settings.components[boost].distribution.parameters.cutoff_frequency_hz
Boost: cutoff (Hz) & $[20,20000]$ & $\mathcal{N}_{\log}(9.2867,0.7411)$ & M (123) \\
% tone.equalization.highshelf.settings.components[boost].distribution.parameters.gain_db
Boost: gain (dB) & $[5,18]$ & $\mathcal{N}(5.0000,2.6958)$ & M (123) \\
% tone.equalization.highshelf.settings.components[boost].distribution.parameters.q
Boost: Q & $[0.1,50]$ & $B_{\log}(11.1019,23.5753)$ & M (123) \\
\addlinespace
\multicolumn{4}{@{}l}{\textbf{Parallel drum compression and distortion}} \\*
% drums.parallel_compression.send_level
Send level & $[0.6,0.85]$ & $U(a, b)$ & $H$ \\
% drums.parallel_compression.return_level
Return level & $[0.3,0.5]$ & $U(a, b)$ & $H$ \\
% drums.parallel_compression.threshold_db
Threshold (dB) & $[-36,-28]$ & $U(a, b)$ & $H$ \\
% drums.parallel_compression.ratio
Ratio & $[6,10]$ & $L(a, b)$ & $H^{\dagger}$ \\
% drums.parallel_compression.attack_ms
Attack (ms) & $[2,10]$ & $L(a, b)$ & $H$ \\
% drums.parallel_compression.release_ms
Release (ms) & $[40,100]$ & $L(a, b)$ & $H$ \\
% drums.distortion.drive_db
Distortion drive (dB) & $[6,14]$ & $U(a, b)$ & $H$ \\
\addlinespace
\multicolumn{4}{@{}l}{\textbf{Limiter}} \\*
% dynamics.limiter.threshold_db
Threshold (dB) & $[-12,-6]$ & $U(a, b)$ & $H$ \\
% dynamics.limiter.release_ms
Release (ms) & $[60,150]$ & $L(a, b)$ & $H$ \\
\addlinespace
\multicolumn{4}{@{}l}{\textbf{Shared ambience: routing, delay, reverb, and filtering}} \\*
% space.shared_send.send_level
Send level & $[0.3,0.8]$ & $U(a, b)$ & $H$ \\
% space.shared_send.return_level
Return level & $[0.6,1.15]$ & $U(a, b)$ & $H$ \\
% space.shared_send.delay.synced.beats
Synced delay (beats) & Listed values & $C\{0.5:2,\ 0.75:3,$\newline $1:1,\ 1.5:0.8\}$ & $H$ \\
% space.shared_send.delay.free.seconds
Free delay (s) & $[0.12,0.75]$ & $U(a, b)$ & $H$ \\
% space.shared_send.delay.mix
Delay mix & $[0.3,0.6]$ & $U(a, b)$ & $H$ \\
% space.shared_send.feedback
Delay feedback & $[0,0.45]$ & $U(a, b)$ & $H$ \\
% space.shared_send.reverb.room_size
Reverb room size & $[0.25,0.95]$ & $U(a, b)$ & $H$ \\
% space.shared_send.reverb.damping
Reverb damping & $[0.25,0.8]$ & $U(a, b)$ & $H$ \\
% space.shared_send.reverb.wet_level
Reverb wet level & $[0.4,0.7]$ & $U(a, b)$ & $H$ \\
% space.shared_send.reverb.dry_level
Reverb dry level & $[0.5,1]$ & $U(a, b)$ & $H$ \\
% space.shared_send.reverb.width
Reverb stereo width & $[0.5,1]$ & $U(a, b)$ & $H$ \\
% space.shared_send.post_eq.highpass_hz
Wet-path high-pass (Hz) & $[700,1000]$ & $L(a, b)$ & $H$ \\
% space.shared_send.post_eq.lowpass_hz
Wet-path low-pass (Hz) & $[2200,3000]$ & $L(a, b)$ & $H$ \\
\addlinespace
\multicolumn{4}{@{}l}{\textbf{Stem balance}} \\*
% balance.target_gain.up_db
Gain increase (dB) & $[3,7]$ & $U(a, b)$ & $H$ \\
% balance.target_gain.down_db
Gain decrease (dB) & $[-7,-3]$ & $U(a, b)$ & $H$ \\
\addlinespace
\multicolumn{4}{@{}l}{\textbf{Cleanup}} \\*
% cleanup.vocal_sibilance.cut_hz
Sibilance cut frequency (Hz) & $[3800,4500]$ & $L(a, b)$ & $H$ \\
% cleanup.vocal_sibilance.cut_db
Sibilance cut gain (dB) & $[-5,-3]$ & $U(a, b)$ & $H$ \\
% cleanup.vocal_sibilance.q
Sibilance Q & $[2.5,4]$ & $L(a, b)$ & $H$ \\
% cleanup.hum.highpass_hz
Hum high-pass (Hz) & $[50,60]$ & $L(a, b)$ & $H$ \\
% cleanup.rumble.highpass_hz
Rumble high-pass (Hz) & $[30,50]$ & $L(a, b)$ & $H$ \\
\addlinespace
\multicolumn{4}{@{}l}{\textbf{Vocal harmony}} \\*
% vocals.harmony.intervals
Pitch shift (semitones) & $\{\pm3,\pm5,\pm7\}$ & $C\{3:4,\ 5:1.5,\ 7:1,$\newline $-3:2,\ {-5}:1.2,\ {-7}:0.8\}$ & $H$ \\
% vocals.harmony.backoff_db
Harmony gain (dB) & $[-6,-1]$ & $U(a, b)$ & $H$ \\
% vocals.harmony.post_eq.highpass_hz
Harmony high-pass (Hz) & $[300,700]$ & $L(a, b)$ & $H$ \\
% vocals.harmony.post_eq.lowpass_hz
Harmony low-pass (Hz) & $[4500,8500]$ & $L(a, b)$ & $H$ \\
\addlinespace
\multicolumn{4}{@{}l}{\textbf{Synthetic noise for cleanup tasks}} \\*
% poison.noise.hum_level_db
Hum level (dB) & $[-30,-24]$ & $U(a, b)$ & $H$ \\
% poison.noise.hum_modulation_hz
Hum modulation (Hz) & $[0.2,0.35]$ & $L(a, b)$ & $H$ \\
% poison.noise.rumble_level_db
Rumble level (dB) & $[-24,-18]$ & $U(a, b)$ & $H$ \\
% poison.noise.rumble_highpass_hz
Rumble high-pass (Hz) & $[18,24]$ & $L(a, b)$ & $H$ \\
% poison.noise.rumble_lowpass_hz
Rumble low-pass (Hz) & $[35,45]$ & $L(a, b)$ & $H$ \\
% poison.noise.rumble_modulation_hz
Rumble modulation (Hz) & $[0.2,0.35]$ & $L(a, b)$ & $H$ \\
\addlinespace
\multicolumn{4}{@{}l}{\textbf{Retro tone}} \\*
% tone.retro.shelf_cut_hz
High-shelf cutoff (Hz) & $[5000,8000]$ & $L(a, b)$ & $H$ \\
% tone.retro.shelf_gain_db
High-shelf gain (dB) & $[-8,-5]$ & $U(a, b)$ & $H$ \\
% tone.retro.lowpass_hz
Low-pass cutoff (Hz) & $[6500,8500]$ & $L(a, b)$ & $H$ \\
\addlinespace
\multicolumn{4}{@{}l}{\textbf{Chorus}} \\*
% tone.modulation.chorus.mix
Mix & $[0.25,0.6]$ & $U(a, b)$ & $H$ \\
% tone.modulation.chorus.depth
Depth & $[0.3,0.65]$ & $U(a, b)$ & $H$ \\
% tone.modulation.chorus.rate_hz
Rate (Hz) & $[0.35,0.9]$ & $L(a, b)$ & $H$ \\
% tone.modulation.chorus.centre_delay_ms
Centre delay (ms) & $[20,50]$ & $U(a, b)$ & $H$ \\
% tone.modulation.chorus.feedback
Feedback & $[0,0.3]$ & $U(a, b)$ & $H$ \\
\end{longtable}

\noindent\textit{Notes.} Parameters without physical units use dimensionless processor controls; delay beats use quarter-note units. For $H$ controls that originally used discrete numerical grids, we use continuous ranges spanning their extrema, except for delay subdivisions and pitch intervals, whose discrete values and weights we retain. $^{\dagger}$We cap the original parallel-compression ratio range at 10 to match our general ratio ceiling. We retain the heuristic retro priors because the eligible observations do not meet our fitting requirements. The configuration contains no fitted peak-EQ prior. The configuration and fit reports retain exact coefficients and contributing source-record locations.

\textbf{Notation key:}
\begin{enumerate}
    \item $L$: log-uniform over $[a,b]$, with density $p(x)=1/[x\ln(b/a)]$. Equal multiplicative intervals receive equal probability.
    \item $U$: uniform over $[a,b]$.
    \item $\mathcal{N}(\mu,\sigma)$: Gaussian, truncated to the listed bounds. $\mu$ and $\sigma$ are its mean and standard deviation before truncation.
    \item $\mathcal{N}_{\log}(\mu,\sigma)$: Gaussian over $\ln x$, truncated to the listed bounds on $x$.
    \item $B(\alpha,\beta)$: beta distribution over the normalized value $(x-a)/(b-a)$.
    \item $B_{\log}(\alpha,\beta)$: beta distribution over $\ln(x/a)/\ln(b/a)$.
    \item $P(\gamma)$: bounded power law, $p(x)\propto x^\gamma$.
    \item $C\{x:w,\ldots\}$: categorical distribution; each value $x$ has probability $w/\sum w$.
    \item $w\mathcal{N}_1 + (1-w)\mathcal{N}_2$: Gaussian mixture; choose component 1 with probability $w$, otherwise 2.
    \item M ($n$), S ($n$): MixParams or SocialFX provenance, with $n$ eligible records.
    \item $H$: heuristic choices made by an audio engineer via listening.
    \item $\dagger$: we capped the parallel-compression ratio at 10.
\end{enumerate}

\endgroup

\subsection{Evaluation Set Recipe Coverage}
\begin{figure}[!htb]
    \centering
    \includegraphics[width=\linewidth]{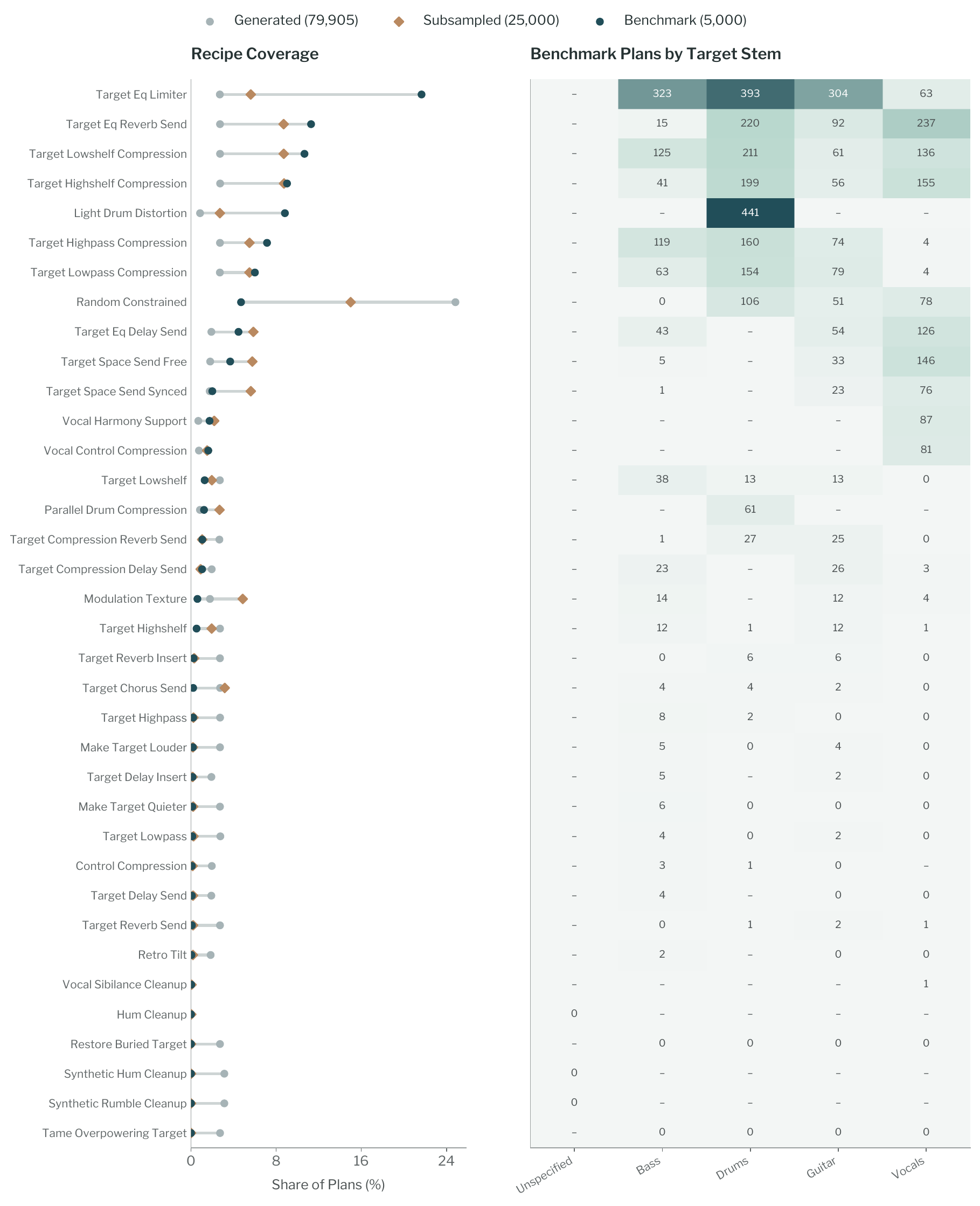}
    \caption{\datasetname recipes in the evaluation dataset, stratified by stem.}
    \label{fig:musiccaps_recipe_coverage_all_stages}
\end{figure}

\Cref{fig:musiccaps_recipe_coverage_all_stages} shows recipe coverage across generation, subsampling, and final benchmark selection. The final 5,000 edits retain 31 of the 32 recipes present in the 25,000 subsampled candidates. The missing recipe, \texttt{hum\_cleanup}, had just six candidates before filtering. Preserving every rare recipe while enforcing instruction validity and edit perceptibility is difficult without explicit coverage quotas, and so we instead prioritize quality criteria over keeping every recipe represented. The recipe imbalance is largely driven by the quality criteria; e.g. high judge acceptance rates and audibility for certain recipes (such as the EQ limiter and EQ reverb send combinations).

\section{Instruction Abstraction Details}
\label{prompts}

\subsection{Abstraction level rules}

We define the following rules for each abstraction level. 

\underline{AL0:}
\begin{itemize}
\item Name every effect explicitly.
\item Name the stem the chain is applied to.
\item State every parameter value exactly as given, with its units.
\item Describe the operators in the same order as the graph.
\item Do not add processing that is not in the graph, and do not leave any out.
\end{itemize}

\underline{AL1:}
\begin{itemize}
\item Name every effect, either by its usual name or by a common word for the thing it does: "grit" or "dirt" for distortion, "volume" for gain.
\item Say which part of the mix you are treating.
\item Do not state any numeric values or units.
\item Convey each parameter with the descriptor supplied for it. The suggested wording is a suggestion, not a required vocabulary: prefer whichever phrasing fits the character of the whole chain, as long as it keeps the parameter at the magnitude shown.
\item Operator order does not need to be preserved.
\end{itemize}

\underline{AL2:}
\begin{itemize}
\item  Summarize the chain as a whole: the overall sonic character it produces and how strongly it is applied. 
\item Describe the combined result rather than the individual processors. Naming an effect is fine where that is the most natural way to convey the character, but it is never required.
\item Express magnitude in words, not numbers or units.
\item Say which part of the mix it applies to. This is the one concrete detail the instruction keeps.
\item Keep it to a single short instruction.
\end{itemize}

\subsection{Abstraction level system prompts}

\paragraph{AL0:} \texttt{ You are a professional mixing engineer writing up a chain for an assistant
      to patch. You know the signal path and you specify it exactly: every
      processor named, every setting given with its units, in the order it is
      patched. You speak economically, the way one engineer talks to another.
      Write one to three spoken sentences.}

\paragraph{AL1:} \texttt{ You are a musician who records and mixes your own material. You know what
      the common effects are called and you use their names, but you do not work
      in numbers: you say how something should sound and how much of it there
      should be. Your vocabulary is a player's, not a spec sheet's.
      Write one or two spoken sentences.}

\paragraph{AL2:} \texttt{You are a musician describing the sound you want rather than how to get it.
      You talk about the overall character of the result and how far it is
      pushed, in the words you would use about a record you like. Naming an
      effect is fine where that is the natural way to say it, but you are asking
      for an outcome, not dictating a recipe.
      Write a single short sentence.}

\subsection{Parameterization bands}

For each effect, we define value ``bands'' for each parameter as well as associated words used as suggestions to describe the band. We derive these vocabularies from \citet{zheng2016socialfx,stasis2017audio} as well as several forums for music production \footnote{Additional references for effect parameter vocabularies:
\href{https://www.musicradar.com/tuition/tech/the-ultimate-guide-to-effects-eq-485820}{[1]}, \href{https://www.musicradar.com/tuition/tech/the-ultimate-guide-to-effects-delay-457920}{[2]}, \href{https://www.musicradar.com/tuition/tech/the-ultimate-guide-to-effects-reverb-461487}{[3]}, \href{https://www.musicradar.com/news/ultimate-guide-modulation-effects}{[4]},
\href{https://www.musicradar.com/news/modern-eq-masterclass}{[5]}, 
\href{https://www.musicradar.com/tuition/tech/the-ultimate-guide-to-effects-filters-469352}{[6]}, \href{https://www.musicradar.com/tuition/tech/the-ultimate-guide-to-effects-distortion-465538}{[7]}}

This is shown in \Cref{tab:param-bands}

\begingroup
\renewcommand{\arraystretch}{1.2}
\setlength{\dashlinedash}{2pt}
\setlength{\dashlinegap}{2pt}
\begin{longtable}{@{} p{3.0cm} p{2.6cm} p{1.5cm} p{7.1cm} @{}}
\caption{Operator parameter descriptor bands.} \label{tab:param-bands} \\
\toprule
\textbf{Effect} & \textbf{Parameter} & \textbf{Band upper} & \textbf{Suggested terms} \\
\midrule
\endfirsthead

\multicolumn{4}{c}{\tablename\ \thetable{} -- continued from previous page} \\
\toprule
\textbf{Effect} & \textbf{Parameter} & \textbf{Band upper} & \textbf{Suggested terms} \\
\midrule
\endhead

\midrule
\multicolumn{4}{r}{continued on next page} \\
\endfoot

\bottomrule
\endlastfoot

% --- separate_audio ---
\multicolumn{4}{@{}l}{\texttt{separate\_audio.description} -- stem selector, not a perceptual parameter; ignored} \\
\midrule

% --- apply_reverb_effect ---
\multirow[t]{10}{3.0cm}{reverb}
 & \texttt{room\_size} & 0.425 & \textit{tight, small, intimate} \\
 & & 0.6   & \textit{medium-sized, moderate, room-sized} \\
 & & 0.775 & \textit{large, spacious, hall-like, distant} \\
 & & 0.96  & \textit{cavernous, vast, enormous, grand, cathedral-like} \\
\cdashline{2-4}
 & \texttt{damping} & 0.4   & \textit{bright, glassy, ringing} \\
 & & 0.675 & \textit{balanced, natural} \\
 & & 0.81  & \textit{dark, damped, muffled, gloomy} \\
\cdashline{2-4}
 & \texttt{width} & 0.67  & \textit{narrow, centered, focused} \\
 & & 0.84  & \textit{moderately wide, open} \\
 & & 1.01  & \textit{wide, full-width, spread right across the stereo field} \\
\addlinespace
 & \multicolumn{3}{l}{\texttt{wet\_level}, \texttt{dry\_level}, \texttt{freeze\_mode} -- pinned/ignored} \\
\midrule

% --- apply_delay_effect ---
\multirow[t]{8}{3.0cm}{delay}
 & \texttt{delay\_seconds} & 0.185 & \textit{slapback, very short} \\
 & & 0.35  & \textit{short, quick echo} \\
 & & 0.6   & \textit{medium, mid-length echo} \\
 & & 3.01  & \textit{long, trailing, spacious echo} \\
\cdashline{2-4}
 & \texttt{feedback} & 0.112 & \textit{a single repeat, no tail} \\
 & & 0.225 & \textit{a couple of repeats, short tail} \\
 & & 0.338 & \textit{several repeats, echoing} \\
 & & 0.46  & \textit{many repeats, regenerating, washed-out tail} \\
\midrule

% --- apply_chorus_effect ---
\multirow[t]{9}{3.0cm}{chorus}
 & \texttt{rate\_hz} & 0.56  & \textit{very slow, drifting, barely moving} \\
 & & 1.01  & \textit{slow, gently swaying} \\
\cdashline{2-4}
 & \texttt{depth} & 0.375 & \textit{mild, gentle, shallow} \\
 & & 0.55  & \textit{moderate, medium} \\
 & & 0.66  & \textit{deep, lush, wide} \\
\cdashline{2-4}
 & \texttt{mix} & 0.325 & \textit{subtle, blended in behind the dry signal} \\
 & & 0.5   & \textit{clearly present, obvious} \\
 & & 0.61  & \textit{heavily blended, prominent, front-and-centre} \\
 & & 1.01  & \textit{fully wet, the whole signal through the effect} \\
\addlinespace
 & \multicolumn{3}{l}{\texttt{centre\_delay\_ms}, \texttt{feedback} -- sampled uniformly; ignored} \\
\midrule

% --- apply_distortion_effect ---
\multirow[t]{3}{3.0cm}{distortion}
 & \texttt{drive\_db} & 8.0   & \textit{light saturation, warmth, a touch of color} \\
 & & 12.0  & \textit{obvious grit, crunch, edge} \\
 & & 15.01 & \textit{heavy, aggressive, crushed} \\
\midrule

% --- apply_compressor_effect ---
\multirow[t]{12}{3.0cm}{compressor}
 & \texttt{ratio} & 3.0   & \textit{gentle, light leveling} \\
 & & 6.0   & \textit{moderate control} \\
 & & 8.0   & \textit{firm, tight} \\
 & & 10.01 & \textit{heavy, squashed, slammed} \\
\cdashline{2-4}
 & \texttt{threshold\_db} & -33.0 & \textit{clamping down on nearly everything} \\
 & & -29.0 & \textit{working on most of the signal} \\
 & & -26.0 & \textit{working on the louder material} \\
 & & 0.01  & \textit{catching only the peaks} \\
\cdashline{2-4}
 & \texttt{attack\_ms} & 6.5   & \textit{fast attack, grabbing transients} \\
 & & 20.01 & \textit{moderate attack, letting transients through} \\
\cdashline{2-4}
 & \texttt{release\_ms} & 58.3  & \textit{quick release, lively, pumping} \\
 & & 200.01 & \textit{slow release, smooth, unobtrusive} \\
\midrule

% --- apply_limiter_effect ---
\multirow[t]{5}{3.0cm}{limiter}
 & \texttt{threshold\_db} & -10.0 & \textit{clamped hard, well into the signal} \\
 & & -8.0  & \textit{holding down the louder peaks} \\
 & & 0.01  & \textit{just shaving the peaks} \\
\cdashline{2-4}
 & \texttt{release\_ms} & 94.9  & \textit{quick release, letting go fast} \\
 & & 200.01 & \textit{slow release, holding the level down} \\
\midrule

% --- apply_gain ---
\multirow[t]{6}{3.0cm}{gain}
 & \texttt{gain\_db} (default) & 4.0   & \textit{a modest amount, noticeably} \\
 & & 6.0   & \textit{substantially, a good deal} \\
 & & 8.01  & \textit{drastically, way up or way down} \\
\cdashline{2-4}
 & \texttt{gain\_db} (\texttt{tuck\_harmony}) & 2.5   & \textit{tucked just under the lead, close behind} \\
 & & 4.5   & \textit{tucked clearly under the lead} \\
 & & 6.01  & \textit{tucked well under the lead, kept in the background} \\
\midrule

% --- shared frequency regions (used by several filters) ---
\multicolumn{4}{@{}l}{\textbf{Shared frequency regions} (\texttt{bands\_ref: shared.frequency\_regions})} \\
 & \texttt{cutoff}/region & 100.0   & \textit{subsonic, infrasonic, below the audible low end} \\
 & & 600.0   & \textit{low end, bass, body, boomy} \\
 & & 1200.0  & \textit{low mids, boxiness, lower midrange} \\
 & & 3400.0  & \textit{upper mids, presence, forward midrange} \\
 & & 4800.0  & \textit{sibilance, harshness, bite, sharp} \\
 & & 22000.0 & \textit{top end, air, brightness, crisp} \\
\addlinespace
 & \multicolumn{3}{l}{Used by: \texttt{highpass\_filter.cutoff\_frequency\_hz};} \\
 & \multicolumn{3}{l}{\texttt{lowpass\_filter.cutoff\_frequency\_hz};} \\
 & \multicolumn{3}{l}{\texttt{highshelf\_filter.cutoff\_frequency\_hz};} \\
 & \multicolumn{3}{l}{\texttt{lowshelf\_filter.cutoff\_frequency\_hz};} \\
 & \multicolumn{3}{l}{\texttt{peak\_filter.cutoff\_frequency\_hz};} \\
 & \multicolumn{3}{l}{\texttt{introduce\_low\_frequency\_rumble.\{highpass\_hz, lowpass\_hz\}}} \\
\midrule

% --- apply_highshelf_filter (gain only; cutoff uses shared regions above) ---
\multirow[t]{4}{3.0cm}{highshelf filter}
 & \texttt{gain\_db} & 6.2   & \textit{a gentle tilt, a slight shelf} \\
 & & 7.4   & \textit{a moderate tilt} \\
 & & 9.4   & \textit{a strong tilt} \\
 & & 18.01 & \textit{a drastic tilt, the top end wholly rebalanced} \\
\midrule

% --- apply_lowshelf_filter (gain, q; cutoff uses shared regions above) ---
\multirow[t]{7}{3.0cm}{lowshelf filter}
 & \texttt{gain\_db} & 2.6   & \textit{a slight tilt, a nudge} \\
 & & 5.5   & \textit{a moderate tilt} \\
 & & 9.5   & \textit{a strong tilt} \\
 & & 60.01 & \textit{a drastic tilt, the low end wholly rebuilt} \\
\cdashline{2-4}
 & \texttt{q} & 0.83  & \textit{a gentle slope, a broad transition} \\
 & & 1.0   & \textit{a standard slope} \\
 & & 50.01 & \textit{a steep slope, a resonant turnover} \\
\midrule

% --- apply_peak_filter (gain, q; cutoff uses shared regions above) ---
\multirow[t]{4}{3.0cm}{peak filter}
 & \texttt{gain\_db} & 4.0   & \textit{a narrow trim, a light dip} \\
 & & 12.01 & \textit{a firm notch, a decisive cut} \\
\cdashline{2-4}
 & \texttt{q} & 3.16  & \textit{moderately narrow, focused} \\
 & & 12.01 & \textit{narrow, surgical, pinpoint} \\
\midrule

% --- introduce_mains_hum ---
\multirow[t]{4}{3.0cm}{mains hum}
 & \texttt{level\_db} & -27.0 & \textit{a faint hum, a low buzz} \\
 & & 0.01  & \textit{an audible hum, an intrusive buzz} \\
\cdashline{2-4}
 & \texttt{modulation\_hz} & 0.265 & \textit{steady, unwavering} \\
 & & 1.01  & \textit{slowly wavering, drifting in level} \\
\addlinespace
 & \multicolumn{3}{l}{\texttt{fundamental\_hz} -- regional mains frequency; ignored} \\
\midrule

% --- introduce_low_frequency_rumble ---
\multirow[t]{4}{3.0cm}{low frequency rumble}
 & \texttt{level\_db} & -21.0 & \textit{a low-level rumble, a faint floor} \\
 & & 0.01  & \textit{a heavy rumble, an intrusive low roar} \\
\cdashline{2-4}
 & \texttt{modulation\_hz} & 0.265 & \textit{steady, unwavering} \\
 & & 1.01  & \textit{slowly wavering, breathing} \\
\addlinespace
 & \multicolumn{3}{l}{\texttt{highpass\_hz}, \texttt{lowpass\_hz} -- see shared frequency regions above} \\
\midrule

% --- send_return block bands (declared on the block, not on an operator) ---
\multicolumn{4}{@{}l}{\textbf{Block bands} (\texttt{block\_bands: send\_return})} \\
\multirow[t]{8}{3.0cm}{send/return}
 & \texttt{send\_level} & 0.4   & \textit{a light feed into the effect} \\
 & & 0.65  & \textit{a moderate feed into the effect} \\
 & & 0.9   & \textit{a strong feed into the effect} \\
 & & 1.01  & \textit{the whole signal fed into the effect} \\
\cdashline{2-4}
 & \texttt{return\_level} & 0.55  & \textit{tucked well under the dry signal, subtle} \\
 & & 0.725 & \textit{sitting under the dry signal, supportive} \\
 & & 1.01  & \textit{clearly audible, level with the dry signal} \\
 & & 1.2   & \textit{pushed above the dry signal, dominant} \\
\addlinespace
 & \multicolumn{3}{l}{\texttt{dry\_level} -- pinned to 1.0 by every recipe; ignored} \\

\end{longtable}
\endgroup

\section{LLM Judge Framework}
\label{app:judge}
\subsection{Prompts}

\paragraph{System Prompt} \begin{quote} 
\small\ttfamily\raggedright\noindent
You are auditing one instruction written for a music-editing dataset.
        You are given the symbolic edit graph it describes, the instruction one
        level of abstraction below it, the rules that instruction was written
        under, and what this level is licensed to discard. You decide two
        things and nothing else. First, whether the instruction obeys its own
        rules. Second, and separately, whether it correctly describes the
        instruction it was derived from -- the same edit, at less precision,
        still truthfully. You are indifferent to style. You do not rewrite
        anything.
        \end{quote}

\paragraph{Rules}
\begin{quote}
\small\ttfamily\raggedright
\begin{enumerate}
\setlength{\itemsep}{4pt}
\setlength{\parskip}{0pt}

\item Judge the instruction only against the rules and contract supplied. Do not
import preferences of your own about how it should read.

\item The contract says what this level MAY discard. Discarding something the
contract permits is never an inconsistency, however much detail is lost. Adding
processing the source does not describe always is.

\item For a banded level, a descriptor is correct when it lands in the band the
graph specifies. The suggested wording is a suggestion: different words in the
right band are correct, and the suggested words in the wrong band are not.

\item For consistency, judge fidelity to the source, not conformance to a
contract. Ask only whether a reader of the rewrite would come away with a true
picture of the edit the source describes. Classify each way it fails:

  \begin{itemize}
  \setlength{\itemsep}{3pt}

  \item wrong\_interpretation \textendash{} the element IS in the source but the
  rewrite describes it incorrectly: a cut described as a boost, a large space
  described as subtle, the wrong stem, one processor mistaken for another. The
  information is present and wrong. Calling a reverb a delay is this, not
  hallucination.

  \item hallucination \textendash{} the rewrite describes processing, a parameter
  or a target that is not in the source at all. Something is ADDED while the rest
  stays intact.

  \item omission \textendash{} something the level was obliged to keep is
  missing. Being less detailed than the source is NOT omission where the licence
  above permits it; that is the abstraction working as intended.

  \end{itemize}

\item A rewrite can fail in more than one way, so list every kind that applies.
An empty list means it describes the source correctly.

\item Do not report the same defect under both verdicts. al\_rules grades the
text against its own level's rubric; consistency grades it against the source. A
missing operator can be either or both, but say so once under each only if it
genuinely fails both tests.

\item Cite rule failures by the index of the rubric item. An unattributable
failure is not a failure.

\item Mechanical checks have already run separately. Judge only what they cannot
settle.

\item Give each failing verdict exactly one sentence of justification, naming the
rubric item or contract term it breaks. The two verdicts get their own sentence
each: they answer different questions and can disagree, so one sentence covering
both explains neither. A passing verdict needs none.

\end{enumerate}
\end{quote}

\subsection{Rejection Reasons}
For the evaluation set, the LLM judge rejected $7,908/23,243$ of the valid subsampled plans (as noted in ~\Cref{sec:experiments}, $1,757$ were rejected prior to filtering). \Cref{tab:reject_reasons} provides a breakdown of these rejections and the corresponding rejection categories provided by the judge. 

\begin{table}[h]
\centering
\begin{tabular}{lr}
\hline
\textbf{Rejection reason} & \textbf{Count} \\
\hline
\texttt{[stylistic]} & 3353 \\
\texttt{[al\_consistency]} & 1717 \\
\texttt{[al\_consistency, stylistic]} & 1205 \\
\texttt{[al\_consistency, al\_rules]} & 912 \\
\texttt{[al\_consistency, al\_rules, stylistic]} & 703 \\
\texttt{[al\_rules]} & 13 \\
\texttt{[al\_rules, stylistic]} & 5 \\
\hline
\end{tabular}

\caption{Evaluation set rejection reasons \& counts.}
\label{tab:reject_reasons}
\end{table}

For the train set, the LLM judge rejected $4,856/25,000$ of the subsampled plans. \Cref{tab:reject_reasons_train} provides a breakdown of these rejections and the corresponding rejection categories provided by the judge. An additional $46$ plans were excluded as a result of Nemotron being unable to generate a valid prompt. Note: since we are not using any captions or real ground-truth data for SFT plans, \texttt{stylistic} was not a criterion for rejection. The criterion \texttt{implementation\_leak} is a regex-based rule that ensures exact MCP tool calls such as \texttt{separate\_audio} or \texttt{apply\_chorus\_effect} do not make it into the instructions verbatim.

\begin{table}[h]
\centering
\begin{tabular}{lr}
\hline
\textbf{Rejection reason} & \textbf{Count} \\
\hline
\texttt{[al\_consistency]} & 2995 \\
\texttt{[al\_consistency, al\_rules]} & 1844 \\
\texttt{[al\_rules]} & 15 \\
\texttt{[al\_consistency, al\_rules, implementation\_leak]} & 2 \\
\hline
\end{tabular}

\caption{SFT Training set rejection reasons \& counts.}
\label{tab:reject_reasons_train}
\end{table}

\section{Agent Framework Details}
\label{app:agent_details}

\subsection{Agent States}

We keep track of four main states in the agent: 

\begin{enumerate}
\item Working Audio State - The agent keeps track of the working audio, the active isolated source, the path of the stem, the path of the residual, and the latest version of the processed stem. 
\item Planning/Assignment State - The tools selected during the planning step are stored in a list of 'assignments.' Each assignment tracks status (pending, in progress, retrying, done, failed) as well as the next pending assignment. 
\item  Execution State - Tracks all aspects of a tool-calling step including current assignment, active source, if a branch (see \Cref{enforcing}) is open, if there are any remaining assignments, and if the audio is ready to return
\item Failure/Repair State - Tracks the failed tool name, failed tool assignment, and failed arguments to the tool call. 
\end{enumerate}

\subsection{Enforcing Agent Behavior }
\label{enforcing}

We introduce a few mechanisms to steer the behavior of the agent:

\begin{itemize}
    \item Separation and Mixing ``Branches'' - When a stem is isolated, a branch is opened and kept track of in the execution state and closed when the source is mixed back in with the residual. This branch keeps track of which source is isolated and the location of the corresponding residual track. The system will not allow another separation tool call or operation on another source until the source has been mixed in with the residual. 
    \item Forced Separation - The system can force separation if the agent tries to operate on a source before isolating that source
    \item Branch Recombination Gate - If a branch is open and there is no pending assignment, the system can force a mix tool call so the agent does not stray
    \item Forced Pending Assignment Processing - If the agent tries to return or mix sources before it has completed all of its assignments, the system will force it to perform another tool call.
    \item Forced return - If there are no more assignments and no branch is open, the system will force the agent to return the audio
\end{itemize}

\subsection{Instrument Priors}

We show the distributions of individual instruments, as well as the number of unique instrument combinations (for only \datasetname-compatible stems) seen in the two reference datasets used to derive priors for the SFT training set in \Cref{tab:instrument_priors}.

\begin{table}[h]
\centering
\small
\setlength{\tabcolsep}{5pt}
\begin{tabular}{@{}lcccccc@{}}
\toprule
\textbf{Source} & \textbf{N} & \textbf{Vocals} & \textbf{Drums} & \textbf{Bass} & \textbf{Guitar} &
\begin{tabular}[b]{@{}c@{}}\textbf{\# unique}\\\textbf{combinations}\end{tabular} \\
\midrule
\citet{nunes2014like} & 2,399 & 96\% & 96\% & 88\% & 39--79\% & not reported \\
MoisesDB~\citep{pereira2023moisesdb} & 240 & 99.6\% & 99.2\% & 98.3\% & 92.5\% & 5/15 \\
\bottomrule
\end{tabular}
\caption{Instrument distribution by source. For \citet{nunes2014like}, guitar representation is given as a range (39--79\%) because separate statistics are reported for ``distorted guitar'', ``acoustic guitar'', and ``clean guitar'', and their overlap is not reported.}
\label{tab:instrument_priors}
\end{table}

\section{\datasetname Validation Interfaces}
\label{app:validation}

We introduce three user interfaces for interactive validation: an edit graph plan generation/subsampling interface, an instruction generation interface, and an interface to test different parameter ranges for \datasetname effects. Instructions to run all interfaces are included in the \datasetname repository.

\subsection{Edit Graph UI}
\begin{figure}[!htb]
    \centering
    \includegraphics[width=\linewidth]{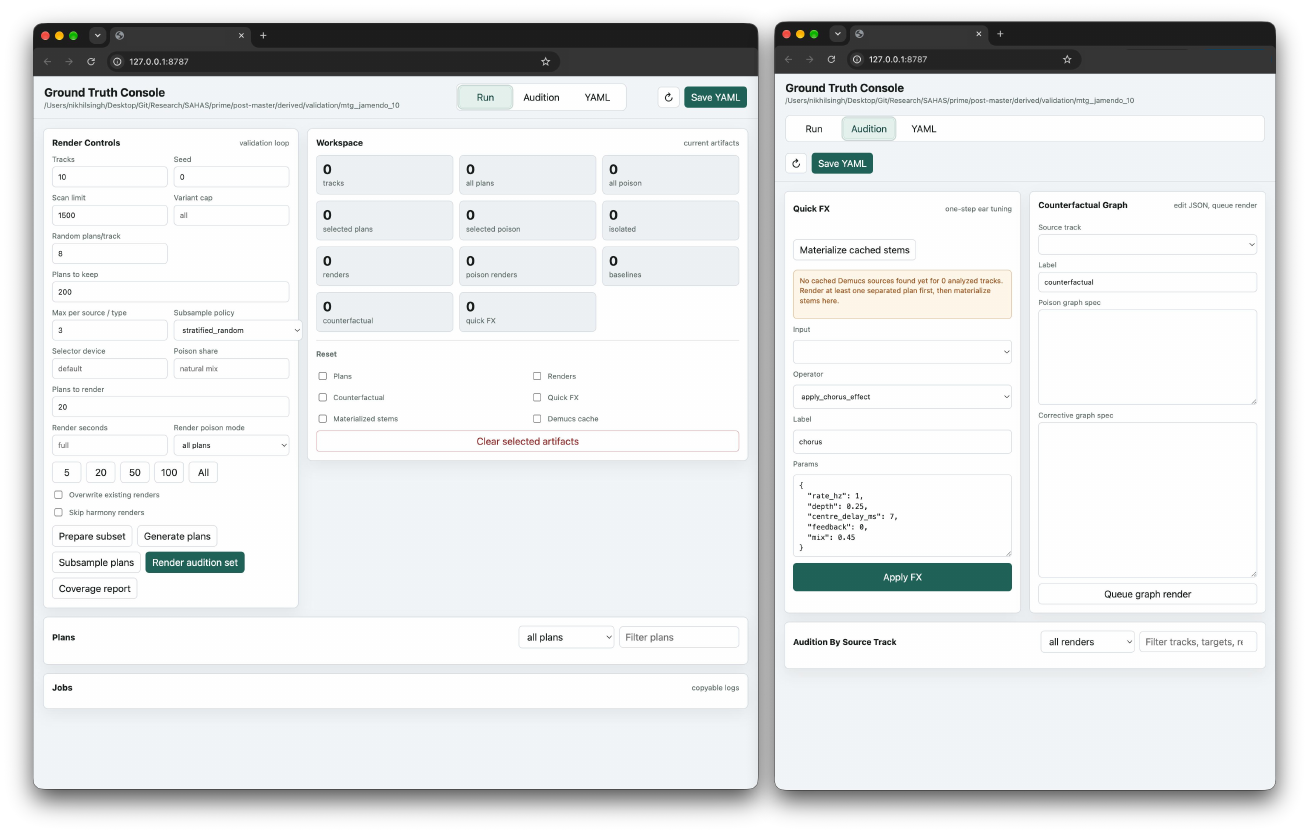}
    \caption{User interface for edit graph generation/subsampling. This was used to validate the procedure outlined in \Cref{alg:graph_gen} before generating plans at scale.}
    \label{fig:audit_ui}
\end{figure}

We include an interface to construct edit graphs from input tracks (\Cref{fig:audit_ui}). From tracks, we can generate edit graphs using \datasetname recipes, motifs, distribution priors, and constraints. Those edit graphs can then be applied to the source audio and listened back to validate the edits are realistic. 

\subsection{Instruction Generation UI}
\begin{figure}[!htb]
    \centering
    \includegraphics[width=\linewidth]{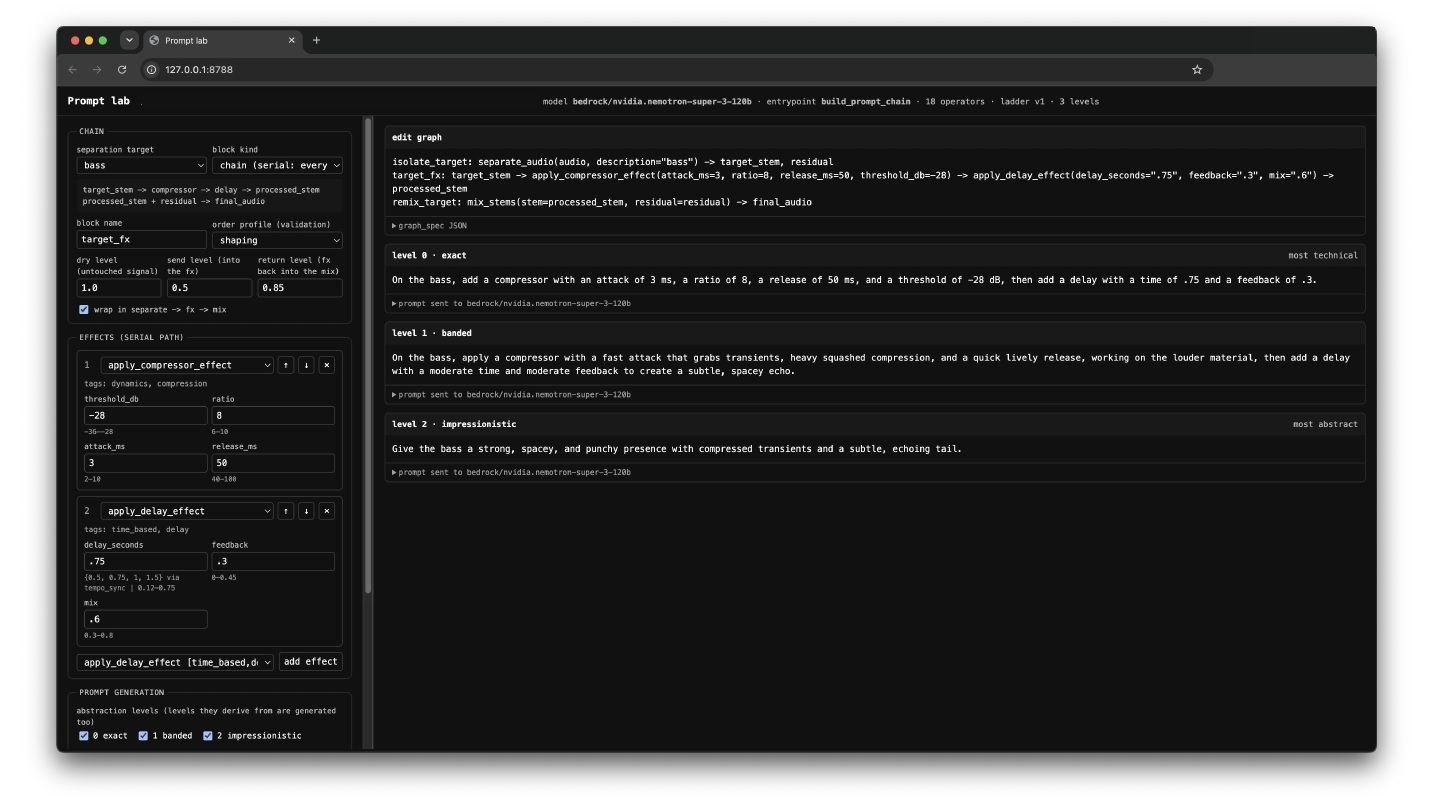}
    \caption{UI for instruction generation. Through this interface, we tested edit graph $\rightarrow$ instruction matching for various language models.}
    \label{fig:prompt_ui}
\end{figure}

The instruction generation UI (shown in \Cref{fig:prompt_ui}) allows you to build an edit graph and specify parameters for each effect and see the resulting instructions generated at different abstraction levels. We use this interface to test different instruction generation models and to assist with prompt engineering. 

\subsection{Parameter Testing UI}
\begin{figure}[!htb]
    \centering
    \includegraphics[width=\linewidth]{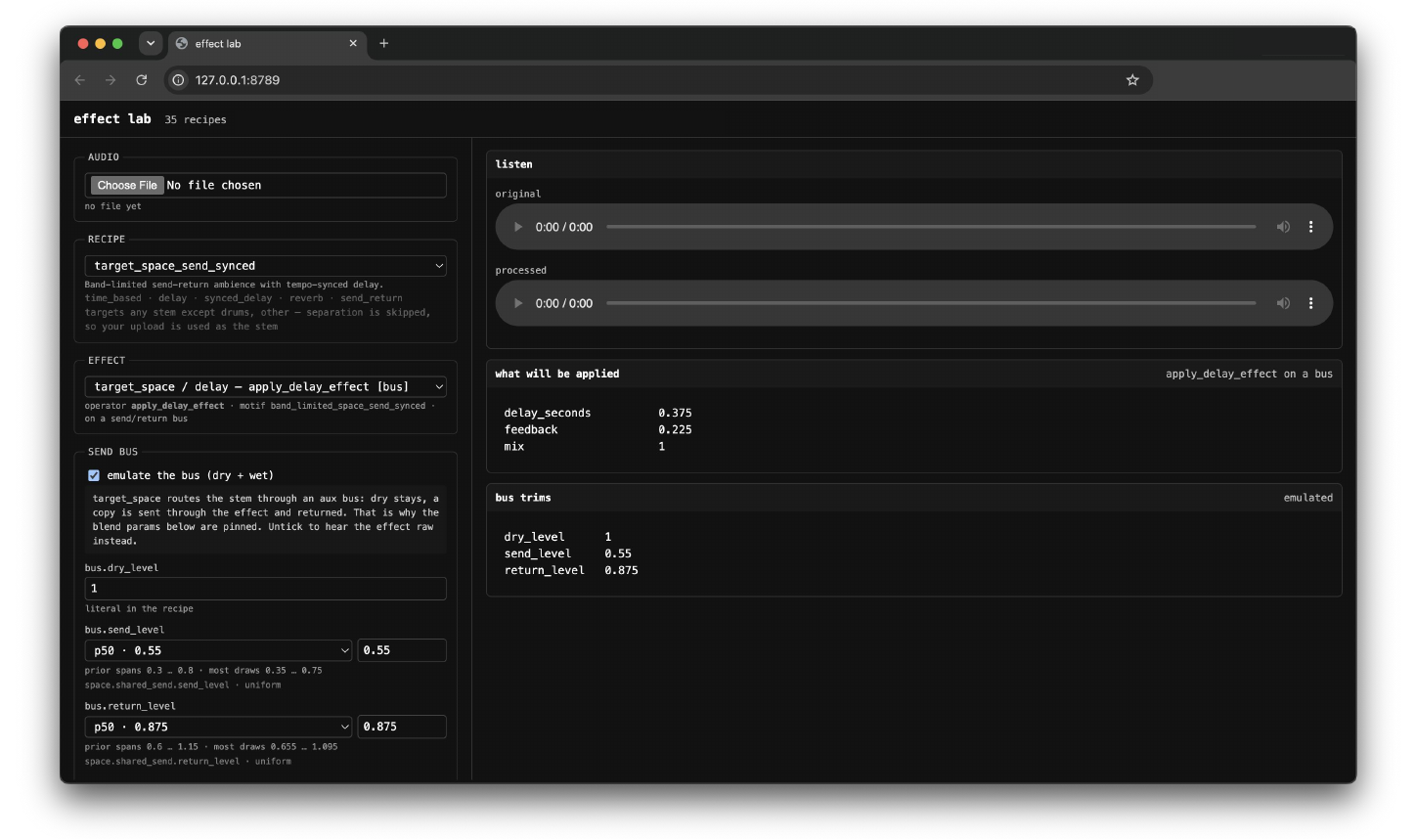} 
    \caption{UI for parameter sampling, used to validate the priors outlined in \Cref{methods:priors} and \Cref{app:priors}.}
    \label{fig:param_ui}
\end{figure}

\Cref{fig:param_ui} shows the parameter testing UI, which lets a user select a specific recipe/effect and sample different values from that effect's priors. A user can then apply that effect with the selected parameters to any input audio source. We used this UI to validate that the derived priors in \Cref{methods:priors} are realistic.

\section{Auxiliary Results}

\subsection{Reference Audio Similarities}
\label{app:aux_results}

\Cref{tab:consolidated_results_full_mert,tab:consolidated_results_full_clap} report mean per-example cosine similarities using pooled MERT and CLAP embeddings, respectively. Input--target similarity measures the size of the reference edit; input--output similarity measures how close the agent stays to the input. Neither reference quantity alone measures task success. Output--target similarity measures absolute audio correspondence, whereas $\Delta\mathrm{sim}_a$ compares the input-relative edit directions. Graph F1 is the mean of kind, stem, operator, and edge F1. All 15,000 attempts per model are included, with failed runs treated as no-ops. CLAP input--target values differ across runs, so cross-model CLAP comparisons require caution.

\begin{table}[h!]
    \centering
    \small
    \caption{Mean reference and output audio similarities using pooled MERT embeddings, together with Graph F1. $N$ counts evaluation attempts.}
    \begin{tabular}{lrrrrrr}
\toprule
Model & $N$ & Input--Target & Input--Output & Output--Target $\uparrow$ & $\Delta \mathrm{sim}_a$ $\uparrow$ & Graph F1 $\uparrow$ \\
\midrule
\gpt & 15000 & 0.972 & 0.971 & 0.979 & 0.637 & 0.821 \\
\gemini & 15000 & 0.972 & 0.971 & 0.971 & 0.528 & 0.634 \\
\gemma & 15000 & 0.972 & 0.958 & 0.965 & 0.609 & 0.857 \\
\gemmasft & 15000 & 0.972 & 0.962 & 0.971 & 0.615 & 0.858 \\
\bottomrule
\end{tabular}

    \label{tab:consolidated_results_full_mert}
\end{table}

\begin{table}[h!]
    \centering
    \small
    \caption{The same reference and output similarities using the CLAP embeddings supplied with each run. Graph F1 is unchanged because it does not use audio embeddings.}
    \begin{tabular}{lrrrrrr}
\toprule
Model & $N$ & Input--Target & Input--Output & Output--Target $\uparrow$ & $\Delta \mathrm{sim}_a$ $\uparrow$ & Graph F1 $\uparrow$ \\
\midrule
\gpt & 15000 & 0.860 & 0.856 & 0.901 & 0.633 & 0.821 \\
\gemini & 15000 & 0.763 & 0.782 & 0.858 & 0.625 & 0.634 \\
\gemma & 15000 & 0.860 & 0.823 & 0.872 & 0.632 & 0.857 \\
\gemmasft & 15000 & 0.763 & 0.728 & 0.877 & 0.730 & 0.858 \\
\bottomrule
\end{tabular}

    \label{tab:consolidated_results_full_clap}

\end{table}

\subsection{Head-to-head Model Performance}
\begin{figure}[!htb]
    \centering
    \includegraphics[width=0.7\linewidth]{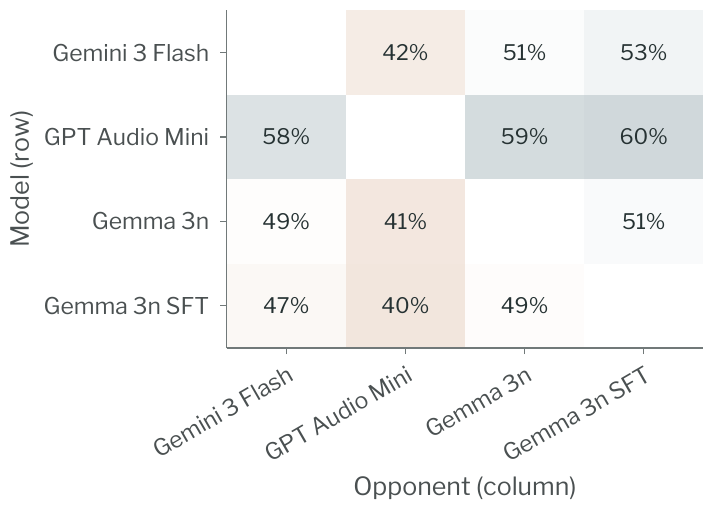}
    \caption{Model pairwise win rates for audio matching.}
    \label{fig:pairwise_wins}
\end{figure}

\begin{figure}[!htb]
    \centering
    \includegraphics[width=\linewidth]{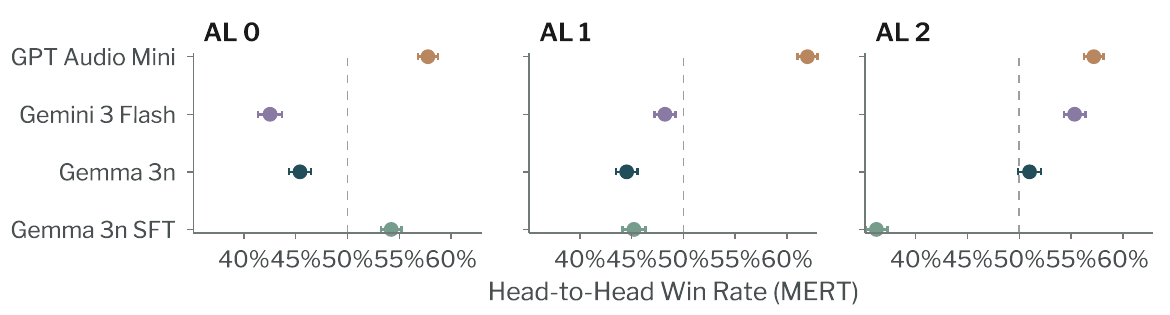}
    \caption{MERT audio-match win rates by abstraction level, averaged across opponents. Each model has 15,000 pairwise comparisons per level. Error bars are nominal 95\% binomial intervals.}
    \label{fig:winrate_aggregate}
\end{figure}

We compare models on the same edit and abstraction level using MERT output--target cosine similarity (\Cref{fig:winrate_aggregate,fig:pairwise_wins}). A win requires strictly higher similarity; ties count as wins for neither model. \gpt has the highest aggregate win rate, $59.0\%$, averaged across the three opponents and abstraction levels. At \alz, \gemmasft beats both \gemini and \gemma in pairwise comparisons and has an aggregate win rate of $54.2\%$. Its aggregate rate drops to $45.2\%$ at \alo and $36.2\%$ at \alt, where it is the lowest of the four models.

\subsection{Model Performance by Graph Complexity}

\begin{figure}[!htb]
\centering
\includegraphics[width=\linewidth]{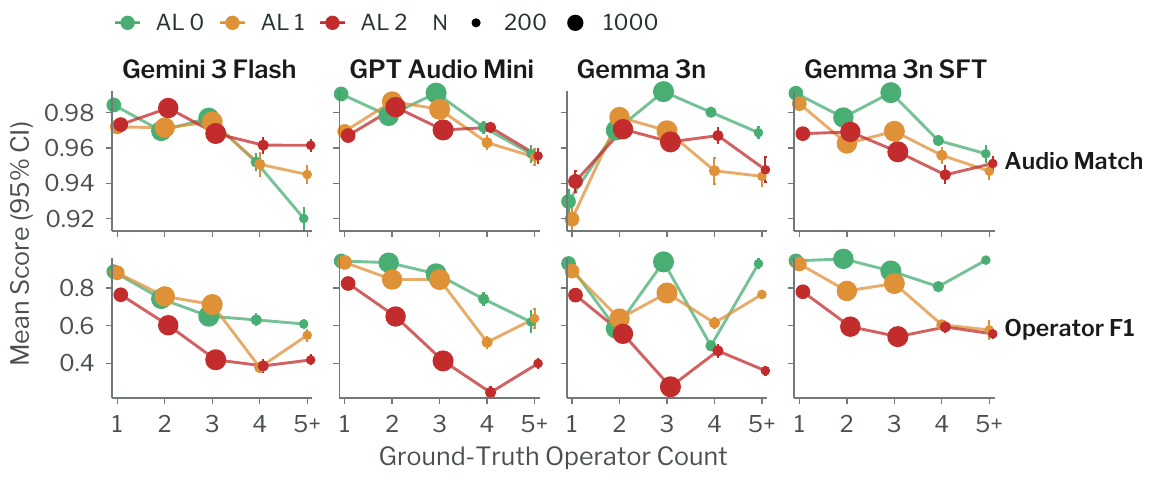}
\vspace{-1.5em}
\caption{MERT output--target audio similarity and operator F1 by ground-truth operator count. Error bars show 95\% CIs.}
\label{fig:complexity_stratification}
\end{figure}

\Cref{fig:complexity_stratification} shows an expanded view of \Cref{fig:complexity_stratification_overall}, stratified by each model. All models show performance degradation as edit graphs get more complex, although this is not monotonic. Generally, they perform worse on the longest edit graphs at higher abstraction levels (though this is not true for \gemini audio match). While fine-tuning seems to help recover more complex graphs at \alz, it seems to have limited effect on \alo and \alt.

\subsection{Further Abstraction Level Analysis}

\begin{figure}[!htb]
    \centering
    \includegraphics[width=\linewidth]{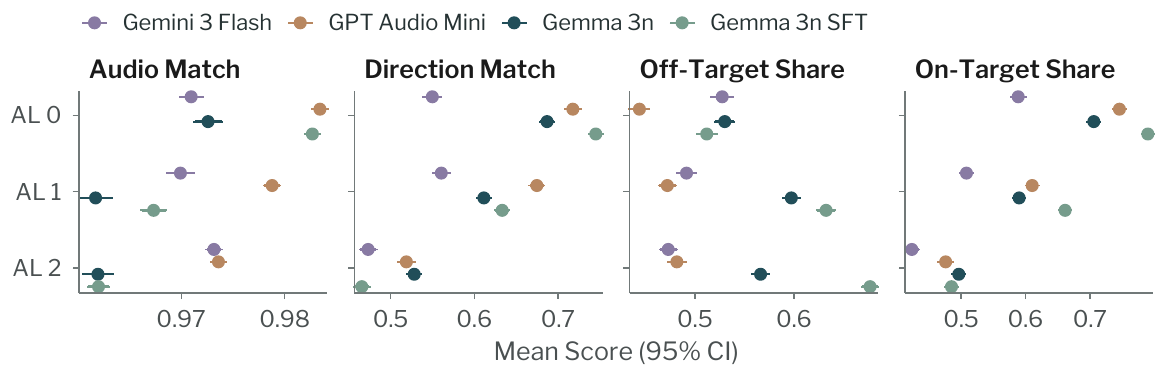}
    \caption{Audio and graph abstraction level metrics.}
    \label{fig:profiles}
\end{figure}

\begin{figure}[!htb]
    \centering
    \includegraphics[width=\linewidth]{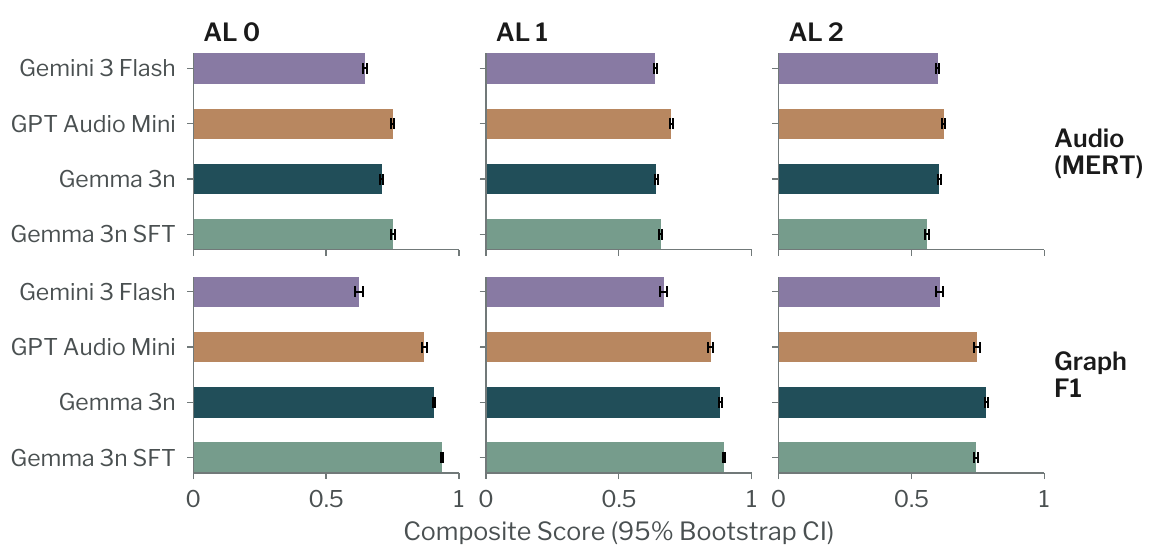}
    \caption{Audio and graph composite scores with confidence intervals by abstraction level.}
    \label{fig:leaderboard_bars_ci}
\end{figure}

\begin{figure}[!htb]
    \centering
    \includegraphics[width=\linewidth]{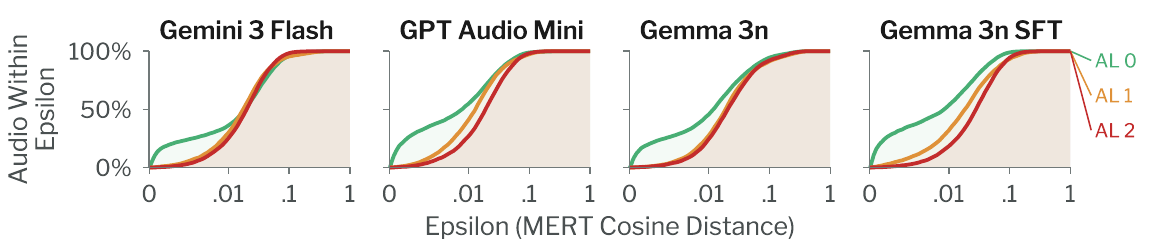}
    \caption{Epsilon thresholding accuracy by abstraction level.}
    \label{fig:epsilon_ball}
\end{figure}

We further explore the role of abstraction level in agent performance. \Cref{fig:profiles} highlights a clear decrease in both audio match and direction match for all models except \gemini from \alz to \alt. \Cref{fig:leaderboard_bars_ci} shows a decrease by abstraction level in audio composite and graph composite scores for both \gpt, \gemma, and \gemmasft. \gemmasft outperforms all models in audio and graph match at \alz and is comparable at \alo, although it falls behind substantially at \alt. \Cref{fig:epsilon_ball} shows that all models perform well at low $\epsilon$ values ($<0.01$) only at \alz, and have almost no audio within tolerance values lower than $0.01$ at \alo and \alt. \gemmasft indicates that fine-tuning is effective at both \alz and \alo (albeit to a lesser extent) but makes little difference at \alt.

\FloatBarrier
\section{Compute Resources}

All \gemini and \gpt experiments were performed via API calls to the respective provider APIs. \gemma and \gemmasft models were hosted on Google Cloud and called via API. \gemmasft was trained on a H200, and embeddings were computed via a mixture of H200s and L40s.

\end{document}